\documentclass{biometrika}

\pdfoutput=1
\RequirePackage{amsmath}
\RequirePackage{natbib}
\usepackage{times}
\usepackage{bm}
\usepackage{amsfonts}
\usepackage{amssymb}
\usepackage{graphicx}
\usepackage{esint}
\usepackage{eucal} % change the font of \mathcal
\usepackage{bbm} % Need for the indicator function notation
\usepackage{fixltx2e} % Fix the 'textsubscript' command
\usepackage{mathrsfs} % Define the math script font
\usepackage{subfig}
\usepackage[font=small]{caption}

\renewcommand{\pmb}{{}}

\def\ftil{\widetilde{f}}
  
   \def\bh{\pmb{h}} \def\hbh{\widehat{\bh}}

 \def\hq{\widehat{q}} 

\def\bs{\pmb{s}}
\def\bt{\pmb{t}}
 
\def\bv{\pmb{v}}   
\def\bw{\pmb{w}}  
\def\bx{\pmb{x}}  
\def\by{\pmb{y}}

\def\bS{\pmb{S}}
\def\bT{\pmb{T}}
 
\def\bX{\pmb{X}}  
\def\bY{\pmb{Y}}

\def\bmu{\pmb{\mu}}

\def\tpi{\widetilde{\pi}}

\def\btheta{\pmb{\theta}}
\def\htheta{\widehat{\theta}}

\def\argmax{\mbox{argmax}}

\newcommand{\tsub}{\textsubscript}

\begin{document}

% \jname{}
% %% The year, volume, and number are determined on publication
% \jyear{}
% \jvol{}
% \jnum{}
% %% The \doi{...} and \accessdate commands are used by the production team
% %\doi{10.1093/biomet/asm023}
% \accessdate{}

% %% These dates are usually set by the production team
% \received{}
% \revised{}

%% The left and right page headers are defined here:
\markboth{W. Li and P. Fearnhead}{Approximate Bayesian Computation asymptotics}

%% Here are the title, author names and addresses
\title{On the Asymptotic Efficiency of Approximate Bayesian Computation Estimators}

\author{Wentao Li}
\affil{School of Mathematics, Statistics and Physics,  Newcastle University,
Newcastle upon Tyne
NE1 7RU, 
U.K.
\email{wentao.li@newcastle.ac.uk}}
\author{and Paul Fearnhead}
\affil{Department of Mathematics and Statistics, Lancaster University, Lancaster LA1 4YF, U.K.  \email{p.fearnhead@lancaster.ac.uk}}
\maketitle

\begin{abstract}
Many statistical applications involve models for which it is difficult to evaluate the likelihood, but from which it is relatively easy to sample.  
Approximate Bayesian computation is a likelihood-free method for implementing Bayesian inference in such cases. We present results on the asymptotic variance of estimators obtained using approximate
Bayesian computation in a large-data limit. Our key assumption is that the data are summarized by a fixed-dimensional summary statistic
that obeys a central limit theorem. We prove asymptotic normality of the mean of the approximate Bayesian computation posterior. 
This result also shows that, in terms of asymptotic variance, we should use a summary statistic that
is the same dimension as the parameter vector, $p$; and that any summary statistic of higher dimension can be reduced, through a linear transformation, to dimension $p$ in a way that can only
reduce the asymptotic variance of the posterior mean. We look at how the Monte Carlo error of an importance sampling algorithm that samples from the approximate Bayesian computation posterior affects the accuracy
of estimators. We give conditions on the importance sampling proposal distribution such that the variance of the estimator will be the 
same order as that of the maximum likelihood estimator based on the
summary statistics used. This  suggests an iterative importance sampling algorithm, which we evaluate empirically on a stochastic volatility model.
\end{abstract}


\section{Introduction}

Many statistical applications involve inference about
models that are easy to simulate from, but for which it is difficult,
or impossible, to calculate likelihoods. In such situations it is
possible to use the fact we can simulate from the model to enable
us to perform inference. There is a wide class of such likelihood-free
methods of inference including indirect inference \cite[]{Gourieroux:1993,Frigessi:2004},
the bootstrap filter \cite[]{Gordon/Salmond/Smith:1993}, simulated
methods of moments \cite[]{duffie1993simulated}, and synthetic likelihood
\cite[]{wood2010statistical}.

We consider a Bayesian version of these methods, termed approximate
Bayesian computation. This involves defining an approximation
to the posterior distribution in such a way that it is possible to
sample from this approximate posterior using only the ability to sample
from the model. Arguably the first approximate
Bayesian computation method was that of \cite{Pritchard:1999},
and these methods have been popular within population genetics \cite[]{Beaumont:2002},
ecology \cite[]{beaumont2010approximate} and systems biology \cite[]{toni2009approximate}.
More recently, there have been applications  to areas
including stereology \cite[]{bortot2007inference},  finance \cite[]{peters2011bayesian}
and cosmology \cite[]{ishida2015cosmoabc}.

Let $K(\bx)$ be a density kernel, scaled, without loss of generality,
so that $\max_{\bx}K(\bx)=1$. 
Further, let $\varepsilon>0$ be a bandwidth. Denote the data by $\bY_{\rm obs}=(y_{\rm obs,1},\ldots,y_{\rm obs,n})$.
Assume we have chosen a finite-dimensional summary statistic $\bs_{n}(\bY)$,
and denote $\bs_{\rm obs}=\bs_{n}(\bY_{\rm obs})$. If we model the data as
a draw from a parametric density, $f_{n}(\by\mid\btheta)$, and assume
prior, $\pi(\btheta)$, then we define the approximate Bayesian computation posterior as 
\begin{equation}
\pi_{\rm ABC}(\btheta\mid \bs_{\rm obs},\varepsilon)\propto\pi(\btheta)\int f_{n}(\bs_{\rm obs}+\varepsilon\bv\mid \btheta)K(\bv)\,d\bv,\label{eq:piABC}
\end{equation}
where $f_{n}(\bs\mid \btheta)$ is the density for the summary statistic
implied by $f_{n}(\by\mid \btheta)$. Let $f_{\rm ABC}(\bs_{\rm obs}\mid \btheta,\varepsilon)=\int f_{n}(\bs_{\rm obs}+\varepsilon\bv\mid \btheta)K(\bv)\,d\bv$.
This framework encompasses most implementations of approximate Bayesian computation. In particular,
the use of the uniform kernel corresponds to the popular rejection-based
rule \cite[]{Beaumont:2002}.

The idea is that $f_{\rm ABC}(\bs_{\rm obs}\mid \btheta,\varepsilon)$ is an approximation
of the likelihood. The approximate Bayesian computation posterior, which is proportional to the
prior multiplied by this likelihood approximation, is an approximation
of the true posterior. The likelihood approximation can be interpreted
as a measure of how close, on average, the summary, $\bs_{n}$, simulated
from the model is to the summary for the observed data, $\bs_{\rm obs}$.
The choices of kernel and bandwidth determine the definition of closeness.

By defining the approximate posterior in this way, we can simulate
samples from it using standard Monte Carlo methods. One approach,
that we will focus on later, uses importance sampling. Let $K_{\varepsilon}(\bx)=K(\bx/\varepsilon)$.
Given a proposal density, $q_{n}(\btheta)$, a bandwidth, $\varepsilon$,
and a Monte Carlo sample size, $N$, an importance sampler
would proceed as in Algorithm 1. The set of accepted parameters and
their associated weights provides a Monte Carlo approximation to $\pi_{\rm ABC}$.
If we set $q_{n}(\btheta)=\pi(\btheta)$ then this is just
a rejection sampler. In practice sequential importance
sampling methods are often used to learn a good proposal distribution
\cite[]{beaumont2009adaptive}.

\begin{algo}
{Importance and rejection sampling approximate Bayesian computation \label{alg:ISABC}} 
\begin{tabbing}
   \quad 1. Simulate $\btheta_{1},\ldots,\btheta_{N}\sim q_{n}(\btheta)$; \\
   \quad 2. For each $i=1,\ldots,N$, simulate $\bY^{(i)}=\{y_{1}^{(i)},\ldots,y_{n}^{(i)}\}\sim f_{n}(y\mid \btheta_{i})$;\\
   \quad 3. For each $i=1,\ldots,N$, accept $\btheta_{i}$ with probability $K_{\varepsilon}\{\bs_{n}^{(i)}-\bs_{\rm obs}\}$, where $\bs_{n}^{(i)}=\bs_{n}\{\bY^{(i)}\}$;\\ \quad \quad and define the associated weight as $w_{i}=\pi(\btheta_{i})/q_{n}(\btheta_{i})$.
\end{tabbing}
\end{algo}

There are three choices in implementing approximate Bayesian computation: the choice of summary
statistic, the choice of bandwidth, and the  Monte
Carlo algorithm. For importance sampling, the last of these involves
specifying the Monte Carlo sample size, $N$, and the proposal density,
$q_{n}(\btheta)$. These, roughly, relate to three sources of approximation. To see this, note that as $\varepsilon\rightarrow0$ we would
expect (\ref{eq:piABC}) to converge to the posterior given $\bs_{\rm obs}$
\cite[]{fearnhead2012constructing}. Thus the choice of summary statistic
governs the approximation, or loss of information, between using the
full posterior distribution and using the posterior given the summary.
The value $\varepsilon$ then affects how close the approximate Bayesian computation posterior
is to the posterior given the summary. Finally there is Monte
Carlo error from approximating the approximate Bayesian computation posterior with a Monte
Carlo sample. The Monte Carlo error is not only affected by the  Monte Carlo algorithm, but also by the choices of summary statistic
and bandwidth, which together affect the probability of acceptance
in step 3 of Algorithm \ref{alg:ISABC}. Having a higher-dimensional summary statistic, or a smaller value of $\varepsilon$,
will tend to reduce this acceptance probability and hence increase
the Monte Carlo error. %These three sources of approximation determine the accuracy of the ABC estimator. 

This work studies the interaction between the three sources
of error, when the summary statistics obey a central
limit theorem for large $n$. We are interested
in the efficiency of approximate Baysian computation, where by efficiency we mean that an estimator
obtained from running Algorithm 1 has the same rate of convergence as the maximum likelihood estimator for the parameter given the
summary statistic. In particular, this work is motivated by the question
of whether approximate Bayesian computation can be efficient as $n\rightarrow\infty$ if we have
a fixed Monte Carlo sample size. Intuitively this appears unlikely.
For efficiency we will need $\varepsilon\rightarrow0$ as $n\rightarrow\infty$,
and this corresponds to an increasingly strict condition for acceptance.
Thus we may imagine that the acceptance probability will necessarily
tend to zero as $n$ increases, and we will need an increasing Monte
Carlo sample size to compensate for this.

However our results show that Algorithm \ref{alg:ISABC} can be efficient if we choose proposal distribution, with a suitable scale and location and appropriately heavy
tails. If we use such a proposal distribution and have
a summary statistic of the same dimension as the parameter vector,
then the posterior mean of approximate Bayesian computation is asymptotically unbiased with
a variance that is $1+O(1/N)$ times that of the estimator maximising the likelihood of the summary statistic. This is similar to asymptotic results for indirect inference
\cite[]{Gourieroux:1993,Frigessi:2004}. Our results also lend theoretical support to methods that
choose the bandwidth indirectly by specifying the proportion
of samples that are accepted, as this leads to a bandwidth which
is of the optimal order in $n$.

We first prove a Bernstein-von Mises type theorem
for the posterior mean of  approximate Bayesian computation. This is a non-standard convergence result,
as it is based on the partial information contained in the summary
statistics. For related convergence results see \cite{clarke1995posterior}
and \cite{yuan2004asymptotic}, though these do not
consider the case when the dimension of the summary statistic is larger
than that of the parameter. Dealing with this case introduces extra challenges. 

Our convergence result for the posterior mean of approximate Bayesian computation
has practically important consequences. It shows that
any $d$-dimensional summary with $d>p$ can be projected to a $p$-dimensional summary statistic without any loss of information. Furthermore
it shows that using a summary statistic of dimension $d>p$
can lead to an increased bias, so the asymptotic
variance can be reduced 
if the optimal $p$-dimensional projected summary is used instead. 
If a $d$-dimensional summary is used, with $d>p$, it suggests choosing the
variance of the kernel to match the variance of the summary statistics.

This paper adds to a growing literature on the theoretical
properties of approximate Bayesian computation. 
%Suppose some function of the parameter, $h(\btheta)$, is of interest. 
%Define the bias of approximate Bayesian computation as $\bh_{\rm ABC}-{E}\{\bh(\btheta)\mid \bs_{\rm obs}\}$ where $\bh_{\rm ABC}$
%is the mean of $h(\btheta)$ under the approximate Bayesian computation posterior.
%Initial results focussed on comparing this bias to the Monte Carlo variance of estimating
%$\bh_{\rm ABC}$. 
Initial results focussed on comparing the bias of approximate Bayesian computation to the Monte Carlo error, and how these depend on the choice of $\varepsilon$.
The convergence rate of the bias is shown to be  $O(\varepsilon^{2})$
in various settings \cite[e.g.,][]{barber2015rate}. This
can then be used to consider how the choice of $\varepsilon$
should depend on the Monte Carlo sample size so as to balance
bias and Monte Carlo variability \cite[]{blum2010approximate,barber2015rate,biau2015new}.
There has also been work on consistency of approximate Bayesian computation estimators. \cite{marin2014relevant} consider consistency when performing model choice
and \cite{Frazier:2017} consider consistency for parameter
estimation. The latter work, which appeared after the first version of this paper, includes a
result on the asymptotic normality of the posterior mean similar to our Theorem \ref{thm:1}, albeit under different conditions, and also gives results on the asymptotic form of the posterior obtained using approximate Bayesian
computation. This shows that for many implementations of approximate Bayesian computation, the posterior will over-estimate
the uncertainty in the parameter estimate that it gives. %\todo{Is this OK?}

Finally, a number of papers have looked at the choice of summary statistics
\cite[e.g.,][]{wegmann2009efficient,blum2010approximate,prangle2014semi}.
%Whilst this is not the focus of our paper, 
Our Theorem \ref{thm:1}
gives insight into this. As mentioned above, this result shows
that, in terms of minimising the asymptotic variance, we should use
a summary statistic of the same dimension as the number of parameters.
In particular it supports the suggestion in \cite{fearnhead2012constructing}
of having one summary per parameter, with that summary approximating
the maximum likelihood estimator for that parameter.

\section{Notation and Set-up}

Denote the data by $\bY_{\rm obs}=(y_{{\rm obs},1},\ldots,$ $y_{{\rm obs},n})$,
where $n$ is the sample size, and each observation, $y_{{\rm obs},i}$,
can be of arbitrary dimension. We  make no assumption directly on the data,
but make assumptions on the distribution of the summary statistics.
We consider the asymptotics
as $n\rightarrow\infty$, and denote the density of $\bY_{\rm obs}$
by $f_{n}(\by\mid \btheta)$, where 
$\btheta\in\mathcal{P}\subset\mathbb{R}^p$. We let $\btheta_{0}$ denote the true parameter value,
and $\pi(\btheta)$ its prior distribution.  For a set $A$, let $A^{c}$ be its complement with respect
to the whole space.

We assume that $\btheta_{0}$ is in the interior of the parameter space, and that the prior is differentiable in a neighbourhood of the true parameter: %, as implied by the following condition:
\begin{condition} \label{par_true}
There exists some $\delta_{0}>0$, such
that $\mathcal{P}_{0}=\{\btheta:|\btheta-\btheta_{0}|<\delta_{0}\}\subset\mathcal{P}$, 
%\end{condition}
%
%The differentiability of the prior density around the true parameter will be required. 
%\begin{condition} \label{prior_regular}
$\pi(\btheta)\in C^{1}(\mathcal{P}_{0})$
and $\pi(\btheta_{0})>0$.
\end{condition}

To implement approximate Bayesian computation we will use a $d$-dimensional summary statistic, $\bs_{n}(\bY)\in\mathbb{R}^{d}$;
such as a vector of sample means of appropriately chosen functions.  We assume that $\bs_{n}(\bY)$ has a density function, which depends on $n$, and we denote this by $f_{n}(\bs\mid \btheta)$.
We will use the shorthand $\bS_{n}$ to denote the random variable
with density $f_{n}(\bs\mid \btheta)$. In approximate Bayesian computation we use a kernel, $K(\bx)$,
with $\max_{\bx}K(\bx)=1$, and a bandwidth $\varepsilon>0$. As we
vary $n$ we will often wish to vary $\varepsilon$, and in these
situations we denote the bandwidth by $\varepsilon_{n}$. For Algorithm \ref{alg:ISABC} we require a proposal distribution, $q_{n}(\btheta)$,
and allow this to depend on $n$. We assume the following conditions
on the kernel, which are satisfied by all commonly-used kernels,
\begin{condition} \label{kernel_prop} The kernel satisfies
(i) $\int\bv K(\bv)\,d\bv=0$; (ii) $\int\prod_{k=1}^{l}v_{i_{k}}K(\bv)\,d\bv<\infty$
for any coordinates $(v_{i_{1}},\ldots,v_{i_{l}})$ of $\bv$ and $l\leq p+6$; (iii) $K(\bv)\propto\overline{K}(\|\bv\|_{\Lambda}^{2})$ where $\|\bv\|_{\Lambda}^{2}=\bv^{T}\Lambda\bv$
and $\Lambda$ is a positive-definite matrix, and $K(\bv)$ is a decreasing
function of $\|\bv\|_{\Lambda}$; (iv) $K(\bv)=O(e^{-c_{1}\|\bv\|_{\Lambda}^{\alpha_{1}}})$ for some $\alpha_{1}>0$
and $c_{1}>0$ as $\|\bv\|_{\Lambda}\rightarrow\infty$.
\end{condition}

For a real function $g(\bx)$ denote its $k$th partial
derivative at $\bx=\bx_{0}$ by $D_{x_{k}}g(\bx_{0})$, the gradient
function by $D_{\bx}g(\bx_{0})$ and the Hessian matrix by $H_{\bx}g(\bx_{0})$.
To simplify notation, $D_{\theta_{k}}$, $D_{\btheta}$ and $H_{\btheta}$
are written as $D_{k}$, $D$ and $H$ respectively. For a series
$x_{n}$
we use the notation that for large enough $n$, $x_{n}=\Theta(a_{n})$
if there exist constants $m$ and $M$ such that $0<m<|x_{n}/a_{n}|<M<\infty$,
and $x_{n}=\Omega(a_{n})$ if $|x_{n}/a_{n}|\rightarrow\infty$. For
two square matrices $A$ and $B$, we say $A\leq B$ if $B-A$ is
semi-positive definite, and $A<B$ if $B-A$ is positive definite.

Our theory will focus on estimates of some function, $\bh(\theta)$, of $\theta$, which satisfies
 differentiability and moment conditions that will control the remainder terms in a Taylor-expansions.
\begin{condition} \label{h_Cdifferentiable}
The $k$th coordinate of $\bh(\btheta)$, $h_{k}(\btheta)$, satisfies (i) $h_{k}(\btheta)\in C^{1}(\mathcal{P}_{0})$; (ii) $D_{k}h(\btheta_{0})\neq0$; and (iii)
%\end{condition}
%\begin{condition} \label{h_moments}
%$\int|h_{k}(\btheta)|\pi(\btheta)\,d\btheta<\infty$ and 
$\int h_{k}(\btheta)^{2}\pi(\btheta)\,d\btheta<\infty$. \end{condition}

The asymptotic results presuppose a central limit theorem for the summary
statistic. 
\begin{condition} \label{sum_conv}
There exists a sequence $a_{n}$, with
$a_{n}\rightarrow\infty$ as $n\rightarrow\infty$, a $d$-dimensional
vector $\bs(\btheta)$ and a $d\times d$ matrix $A(\btheta)$, such
that for all $\btheta\in\mathcal{P}_0$, 
\[
a_{n}\{\bS_{n}-\bs(\btheta)\}\rightarrow N\{0,A(\btheta)\}, ~~ \mbox{\ensuremath{n\rightarrow\infty}},
\]
with convergence in distribution. We also assume that $s_{\rm obs}\rightarrow s(\theta_0)$ in probability. Furthermore, (i) $\bs(\btheta)\in C^{1}(\mathcal{P}_{0})$ and $A(\btheta)\in C^{1}(\mathcal{P}_{0})$, and
$A(\btheta)$ is positive definite for $\btheta\in\mathcal{P}_{0}$; (ii) for any $\delta>0$ there exists a $\delta'>0$ such that $\|s(\theta)-s(\theta_0)\|>\delta'$ for all $\theta$ satisfying 
$\|\theta-\theta_0\|>\delta$; (iii) $I(\btheta)=D\bs(\btheta)^{T}A^{-1}(\btheta)D\bs(\btheta)$
has full rank at $\btheta=\btheta_{0}$.
\end{condition}

Under Condition \ref{sum_conv}, $a_{n}$ is the rate
of convergence in the central limit theorem. If the data are independent
and identically distributed, and the summaries are sample means
of functions of the data or of quantiles, then $a_{n}=n^{1/2}$. In most applications the data will be dependent, but if summaries are sample means  \cite[]{wood2010statistical}, 
quantiles \cite[]{peters2011bayesian,Allingham:2008,blum2010non} or linear combinations thereof \cite[]{fearnhead2012constructing} then a central limit theorem will often still hold, though 
%For situations where there is long-range dependence in the data then 
$a_n$ may increase more slowly than $n^{1/2}$.

Part (ii) of
Condition \ref{sum_conv} is required for the true parameter to be identifiable given
only the summary of data. %In the following we will define $I(\theta)=D\bs(\btheta)^{T}A^{-1}(\btheta)D\bs(\btheta)$.
The asymptotic variance of the summary-based maximum likelihood estimator for $\btheta$ is $I^{-1}(\btheta_{0})/a_{n}^{2}$. Condition (iii) ensures that this variance is  valid at the
true parameter.

We next require a condition that controls the difference between $f_{n}(\bs\mid \btheta)$
and its limiting distribution for $\btheta\in\mathcal{P}_{0}$. Let
$N(\bx;\bmu,\Sigma)$ be the normal density at $\bx$ with mean $\bmu$
and variance $\Sigma$. Define $\widetilde{f}_{n}(\bs\mid \btheta)=N\{\bs;\bs(\btheta),A(\btheta)/a_{n}^{2}\}$
and the standardized random variable $W_{n}(\bs)=a_{n}A(\btheta)^{-1/2}\{\bs-\bs(\btheta)\}$.
Let $\widetilde{f}_{W_{n}}(\bw\mid \btheta)$ and $f_{W_{n}}(\bw\mid \btheta)$
be the density of $W_{n}(\bs)$ when $\bs\sim\widetilde{f}_{n}(\bs\mid \btheta)$
and $f_{n}(\bs\mid \btheta)$ respectively. The condition below requires that
the difference between $f_{W_{n}}(\bw\mid \btheta)$ and its Edgeworth
expansion $\widetilde{f}_{W_{n}}(\bw\mid \btheta)$ is $o(a_n^{-2/5})$ and can be bounded by a density with exponentially decreasing tails. 
This is weaker than the standard requirement, $o(a_n^{-1})$, for the remainder in the Edgeworth expansion.
%and can be bounded by a density with exponentially decreasing tails. % Specifically, assume that 

\begin{condition} \label{sum_approx}
There exists $\alpha_{n}$ satisfying $\alpha_{n}/a_n^{2/5}\rightarrow\infty$
and a density $r_{\rm max}(\bw)$ satisfying Condition \ref{kernel_prop} (ii)-(iii) where $K(\bv)$ is replaced with $r_{\rm max}(\bw)$, such that $\sup_{\btheta\in\mathcal{P}_{0}}\alpha_{n}| f_{W_{n}}(\bw\mid \btheta)-\widetilde{f}_{W_{n}}(\bw\mid \btheta)|\leq c_{3}r_{\rm max}(\bw)$
for some positive constant $c_{3}$.
\end{condition}

The following condition further assumes that $f_{n}(s\mid\theta)$ has exponentially decreasing tails 
with rate uniform in the support of $\pi(\theta)$.

\begin{condition} \label{sum_approx_tail}
The following statements hold: (i) $r_{\rm max}(w)$ satisfies Condition \ref{kernel_prop} (iv); 
and 
(ii) $\sup_{\theta\in\mathcal{P}_{0}^{c}}f_{W_{n}}(w\mid\theta)=O(e^{-c_{2}\|w\|^{\alpha_{2}}})$
as $\|w\|\rightarrow\infty$ for some positive constants $c_{2}$
and $\alpha_{2}$, and $A(\theta)$ is bounded in $\mathcal{P}$.
\end{condition}

%\begin{condition} \label{cond:prior_regular}
%It holds that $\pi(\theta)\in C^{2}(\mathcal{P}_{0})$
%and $\pi(\theta_{0})>0$. 
%\end{condition}

\section{Posterior mean asymptotics}

We first ignore any Monte Carlo error, and focus on the ideal estimator of true posterior mean from
approximate Bayesian computation. This is the posterior mean, $\bh_{\rm ABC}$, where 
\[
 \bh_{\rm ABC}={E}_{\pi_{\rm ABC}}\{\bh(\btheta)\mid \bs_{\rm obs}\}= \int \bh(\btheta) \pi_{\rm ABC}(\btheta\mid \bs_{\rm obs},\varepsilon_n). 
\]
This estimator
depends on $\varepsilon_n$, but we suppress this from the notation.
As an approximation to the true posterior mean, ${E}\{\bh(\btheta)\mid \bY_{\rm obs}\}$,
$\bh_{\rm ABC}$ contains errors from the choice of the bandwidth
$\varepsilon_{n}$ and summary statistic $\bs_{\rm obs}$.

To understand the effect of these two sources of error, we derive
results for the asymptotic distributions of $\bh_{\rm ABC}$ and the likelihood-based
estimators, including the summary-based maximum likelihood estimator and the
summary-based posterior mean, where we consider randomness solely
due to the randomness of the data. Let
$T_{{\rm obs}}=a_{n}A(\theta_{0})^{-1/2}\{s_{\rm obs}-s(\theta_{0})\}$.

\begin{theorem} \label{thm:1}Assume Conditions \ref{par_true}--\ref{sum_approx_tail}. 
\begin{itemize}
\item[(i)] Let $\hat{{\btheta}}_{_{\mbox{\scriptsize \rm MLES}}}=\argmax_{\btheta\in\mathcal{P}}\log f_{n}(\bs_{\rm obs}\mid \btheta)$. For $\bh_{\bs}=\bh(\hat{{\btheta}}_{_{\mbox{\scriptsize \rm MLES}}})$
or ${E}\{\bh(\btheta)\mid \bs_{\rm obs}\}$, 
\[a_{n}\{\bh_{\bs}-\bh(\btheta_{0})\}\rightarrow N\{0,D\bh(\btheta_{0})^{T}I^{-1}(\btheta_{0})D\bh(\btheta_{0})\},~~n\rightarrow\infty,\]
with convergence in distribution. 
\item[(ii)] Define $c_{\infty}= \lim_{n\rightarrow\infty}a_{n}\varepsilon_{n}$. Let $Z$ be the weak limit of
$T_{\rm obs}$, which has a standard normal distribution, and $R(c_{\infty},Z)$ be a random vector with mean zero that is defined in the Supplementary Material.
If $\varepsilon_{n}=o(a_{n}^{-3/5})$, then
\[
a_{n}\{h_{{\rm ABC}}-h(\theta_{0})\}\rightarrow Dh(\theta_{0})^{T}\{I(\theta_{0})^{-1/2}Z+R(c_\infty,Z)\}, ~~ n\rightarrow\infty,
\]
with convergence in distribution. If either
 (i) $\varepsilon_{n}=o(a_{n}^{-1})$;
(ii) $d=p$; or %(iii) $A(\theta_{0})$ is diagonal; 
(iii) the covariance
matrix of $K(v)$ is proportional to $A(\theta_{0})$; then $R(c_{\infty},Z)=0$. For other cases,
the variance of $I(\theta_{0})^{-1/2}Z+R(c_{\infty},Z)$ is no less than $I^{-1}(\theta_{0})$. 
\end{itemize}
\end{theorem}
%The explicit form of $R(\lim_{n\rightarrow\infty}a_{n}\varepsilon_{n},Z)$ is given in the supplementary material.

Theorem \ref{thm:1} (i) shows the validity of posterior inference
based on the summary statistics. Regardless of the sufficiency and
dimension of $\bs_{\rm obs}$, the posterior mean based on the summary
statistics is consistent and asymptotically normal with the same variance
as the summary-based maximum likelihood estimator. 

Denote the bias of approximate Bayesian computation, $\bh_{\rm ABC}-{E}\{\bh(\btheta)\mid \bs_{\rm obs}\}$,
by $\mbox{bias}_{\rm ABC}$. The choice of bandwidth impacts the size of the bias. 
Theorem \ref{thm:1} (ii) indicates two regimes for the bandwidth for which the posterior mean of
approximate Bayesian computation has good properties.

The first case is when $\varepsilon_{n}$ is $o(1/a_{n})$. For this regime the
posterior mean of approximate Bayesian computation always has the same
asymptotic distribution as that of the true posterior given the summaries.
The other case is when $\varepsilon_{n}$ is $o(a_n^{-3/5})$ but not $o(n^{-1})$.
We obtain the same asymptotic distribution if either
$d=p$ or we choose the kernel variance to be proportional to the variance of the summary statistics.
In general for this regime of $\varepsilon_n$, $\bh_{\rm ABC}$ will be
less efficient than the summary-based maximum likelihood estimator. %This latter result gives a reason for choosing the dimension of the sufficient statistic to be the same as that of the parameter vector. Whilst this has been argued previously, that was based around a different issue, namely controlling the acceptance rate of Monte Carlo algorithms for sampling from the ABC posterior. %Similar phenomena is observed for indirect inference in \cite{Frigessi:2004}. 
%Although the negligible $\varepsilon_{n}$ is preferred, we show below,
% in Remark \ref{neglegible_bandwidth},with such choices of $\varepsilon_{n}$
%that the Monte Carlo acceptance rate will inevitably degenerate as $n\rightarrow\infty$ for such a regime.

When $d>p$, Theorem \ref{thm:1} (ii) shows that $\mbox{bias}_{\rm ABC}$
is non-negligible and can increase the asymptotic variance. This is
because the leading term of $\mbox{bias}_{\rm ABC}$ is proportional to
the average of $\bv=\bs-\bs_{\rm obs}$, the difference between the simulated
and observed summary statistics. If $d>p$, the marginal density
of $\bv$ is generally asymmetric, and thus is no longer guaranteed to have a mean of zero. 
One way to ensure that there is no increase in the asymptotic variance is to choose the variance of the kernel to
be proportional to the variance of the summary statistics.

The loss of efficiency we observe in Theorem
\ref{thm:1} (ii) for $d>p$ gives an advantage for choosing
a summary statistic with $d=p$. The following proposition
shows that for any summary statistic of dimension $d>p$ we can find
a new $p$-dimensional summary statistic without any loss of information.
The proof of the proposition is trivial and hence omitted.
%How to choose the dimension $d$ of $\bs_{\rm obs}$ is of interest, since larger $d$ gives possibly more informative $\bs_{\rm obs}$ but slower convergence of $\hbh$ when $N$ increases \cite[]{blum2013comparative} and increased variance as indicated in Theorem \ref{thm:1}. From Remark \ref{rem_dimension}, it can be seen that for $\bs_{\rm obs}$, only its projection in the row space of $D\bs(\btheta_{0})^{T}A(\btheta_{0})^{-1/2}$ are effective. Therefore when $d$ exceeds the dimension of the parameter, we can use a $p$ dimensional statistic without loss of information and also remove the extra variation. This is stated in the following proposition.
\begin{proposition}
\label{prop:dim_reduce} Assume the conditions of Theorem \ref{thm:1}.
If $d>p$, define $C=D\bs(\btheta_{0})^{T}A(\btheta_{0})^{-1}$. The $p$-dimensional
summary statistic $C\bS_{n}$ has the same information matrix, $I(\btheta)$, as $\bS_{n}$.
Therefore the asymptotic variance of $\bh_{\rm ABC}$ based on $C\bs_{\rm obs}$
is smaller than or equal to that based on $\bs_{\rm obs}$. \end{proposition}
%\begin{proof} The equality can be verified by algebra. \end{proof}
%The proposition shows that a  linear projection can be an
%effective dimension reduction method, when $\varepsilon_{n}$ is small
%enough that the condition in Theorem \ref{thm:1} (ii) is satisfied.
%The matrix $C$ can be interpreted as the product of the scale matrix
%$A(\btheta_{0})^{-1/2}$, which standardizes $\bs_{\rm obs}$, and the
%matrix $D\bs(\btheta_{0})^{T}A(\btheta_{0})^{-1/2}$ which can be
%taken as the `squared-root' of $I(\btheta_{0})$.

Theorem \ref{thm:1} leads to following natural definition. \begin{definition}
Assume that the conditions of Theorem \ref{thm:1} hold. Then the
{\em asymptotic variance} of $\bh_{\rm ABC}$ is 
\[
\textsc{AV}_{\bh_{\rm ABC}}=\frac{1}{a_{n}^{2}}D\bh(\btheta_{0})^{T}I_{\rm ABC}^{-1}(\btheta_{0})D\bh(\btheta_{0}).
\]
\end{definition}

\section{Asymptotic Properties of Rejection and Importance Sampling Algorithm}

\subsection{Asymptotic Monte Carlo Error}

We now consider the Monte Carlo error involved in estimating $\bh_{\rm ABC}$.
Here we fix the data and consider  solely  the
stochasticity of the Monte Carlo algorithm. We focus on Algorithm \ref{alg:ISABC}. Remember that $N$ is
the Monte Carlo sample size. For $i=1,\ldots,N$, $\btheta_{i}$ is
the proposed parameter value and $w_{i}$ is its importance sampling
weight. Let $\phi_{i}$ be the indicator that is 1 if and only if
$\theta_{i}$ is accepted in step 3 of Algorithm \ref{alg:ISABC} and let $N_{\rm acc}=\sum_{i=1}^{N}\phi_{i}$
be the number of accepted parameter.

Provided $N_{\rm acc}\geq1$ we can estimate $\bh_{\rm ABC}$ from the output
of Algorithm \ref{alg:ISABC} with 
\[
\hbh=\sum_{i=1}^{N}\bh(\btheta_{i})w_{i}\phi_{i}\Big/\sum_{i=1}^{N}w_{i}\phi_{i}.
\]
Define the acceptance probability
\[
p_{{\rm acc},q}=\int q(\btheta)\int f_{n}(\bs\mid \btheta)K_{\varepsilon}(\bs-\bs_{\rm obs})\mbox{d}\bs\mbox{d}\btheta,
\]
and the density of the accepted parameter
\[
q_{\rm ABC}(\btheta\mid \bs_{\rm obs},\varepsilon)=\frac{q_{n}(\btheta)f_{\rm ABC}(\bs_{\rm obs}\mid \btheta,\varepsilon)}{\int q_{n}(\btheta)f_{\rm ABC}(\bs_{\rm obs}\mid \btheta,\varepsilon)\,d\theta}.
\]
Finally, define
\begin{align}
 & \Sigma_{{\rm IS},n}=E_{\pi_{\rm ABC}}\left\{(\bh(\btheta)-\bh_{\rm ABC})^{2}\frac{\pi_{\rm ABC}(\btheta\mid \bs_{\rm obs},\varepsilon_{n})}{q_{\rm ABC}(\btheta\mid \bs_{\rm obs},\varepsilon_{n})}\right\},\nonumber \\
 & \Sigma_{{\rm ABC},n}=p_{{\rm acc},q_{n}}^{-1}\Sigma_{{\rm IS},n},\label{ISABC_var}
\end{align}
where $\Sigma_{{\rm IS},n}$ is the importance sampling variance with $\pi_{\rm ABC}$ as the
target density and $q_{\rm ABC}$ as the proposal density. Note that $p_{{\rm acc},q_{n}}$
and $\Sigma_{{\rm IS},n}$, and hence $\Sigma_{{\rm ABC},n}$, depend on $\bs_{\rm obs}$.

Standard results give the following asymptotic distribution of $\hbh$.
\begin{proposition} \label{prop:Monte_Carlo_CLT} For a given $n$ and $\bs_{\rm obs}$,
if $\bh_{\rm ABC}$ and $\Sigma_{{\rm ABC},n}$ are finite, then 
\[
{N}^{1/2}(\hbh-\bh_{\rm ABC})\rightarrow N(0,\Sigma_{{\rm ABC},n}),
\]
in distribution as $N\rightarrow\infty$. \end{proposition}
This proposition motivates the
following definition. 
\begin{definition} For a given $n$ and $\bs_{\rm obs}$,
assume that the conditions of Proposition \ref{prop:Monte_Carlo_CLT}
hold. Then the  asymptotic Monte Carlo variance of $\hbh$
is 
\[
\textsc{MCV}_{\hbh}=\frac{1}{N}\Sigma_{{\rm ABC},n}.
\]
\end{definition}

%From Proposition \ref{prop:Monte_Carlo_CLT}, it can be seen that
%the asymptotic Monte Carlo variance of $\hbh$ 
%is equal to the importance sampling
%variance $\Sigma_{{\rm IS},n}$ divided by the average number of acceptance
%$Np_{{\rm acc},q_{n}}$, and therefore depends on the proposal distribution
%and $\varepsilon_{n}$ through these two terms.

%\begin{rem}[Optimal proposal density] \label{rem3}%According the alternative expression of $\Sigma_{{\rm ABC},n}$ in the proof of Proposition \ref{prop:Monte_Carlo_CLT} we see that  %the optimal proposal density, minimising $\textsc{MCV}_{\hbh}$, is the density proportional to $\big|\bh(\btheta)-\bh_{\rm ABC}\big|\pi(\btheta)f_{\rm ABC}(\bs_{\rm obs}\mid \btheta,\varepsilon)^{1/2}$. %This can be obtained similarly to obtaining the optimal proposal density for the ratio estimate of importance sampling \cite[Chapter2]{hesterberg1988advances}.%\end{rem}

\subsection{Asymptotic efficiency}

We have defined the asymptotic variance as $n\rightarrow\infty$ of
$\bh_{\rm ABC}$, and the asymptotic Monte Carlo variance, as $N\rightarrow\infty$
of $\hbh$. The error of $\bh_{\rm ABC}$ when estimating $\bh(\btheta_{0})$
and the Monte Carlo error of $\hbh$ when estimating $\bh_{\rm ABC}$
are independent, which suggests the following definition.
\begin{definition} Assume the conditions of Theorem \ref{thm:1},
and that $\bh_{\rm ABC}$ and $\Sigma_{{\rm ABC},n}$ are bounded in probability
for any $n$. Then the {\em asymptotic variance} of $\hbh$ is
\[
\textsc{AV}_{\hbh}=\frac{1}{a_{n}^{2}}\bh(\btheta_{0})^{T}I_{\rm ABC}^{-1}(\btheta_{0})D\bh(\btheta_{0})+\frac{1}{N}\Sigma_{{\rm ABC},n}.
\]
\end{definition} 
We can interpret the asymptotic variance of $\hbh$ as a first-order approximation to the
variance of our Monte Carlo estimator for both large $n$ and $N$. 
We wish to investigate the properties of this asymptotic variance,
for large but fixed $N$, as $n\rightarrow\infty$. The asymptotic variance itself depends on $n$, and we would
hope it would tend to zero as $n$ increases. Thus we will study the ratio of $\textsc{AV}_{\hbh}$ to $\textsc{AV}_{\mbox{\scriptsize \rm MLES}}$,
where, by Theorem \ref{thm:1}, the latter is $a_{n}^{-2}\bh(\btheta_{0})^{T}I^{-1}(\btheta_{0})D\bh(\btheta_{0})$.
This ratio measures the efficiency of our Monte Carlo estimator relative to the maximum likelihood
estimator based on the summaries; it quantifies the loss of efficiency from using a non-zero bandwidth and a finite
Monte Carlo sample size.

We will consider how this ratio depends on the choice of $\varepsilon_{n}$
and $q_{n}(\btheta)$. Thus we introduce the following definition:
\begin{definition} For a choice of $\varepsilon_{n}$ and $q_{n}(\btheta)$,
we define the asymptotic efficiency of $\hbh$ as 
\[
\textsc{AE}_{\hbh}=\lim_{n\rightarrow\infty}\frac{\textsc{AV}_{\mbox{\scriptsize \rm MLES}}}{\textsc{AV}_{\hbh}}.
\]
If this limiting value is zero, we say that $\hbh$ is asymptotically
inefficient. \end{definition}

We will investigate the asymptotic efficiency of $\hbh$ under the
assumption of Theorem \ref{thm:1} that $\varepsilon_{n}=o(a_{n}^{-3/5})$.
We shall see that the convergence rate of the importance sampling variance $\Sigma_{{\rm IS},n}$
depends on how large $\varepsilon_{n}$ is relative to $a_n$, and so we further define
$a_{n,\varepsilon}=a_n$ if $\lim_{n\rightarrow\infty} a_n\varepsilon_n<\infty$ and $a_{n,\varepsilon}=\varepsilon_n^{-1}$ otherwise.

If our proposal distribution in Algorithm \ref{alg:ISABC} is either the prior or the posterior, then the estimator is asymptotically inefficient.
\begin{theorem}\label{thm:2}
Assume the conditions of Theorem \ref{thm:1}. %Consider a fixed $N$. Then
%We have: 
\begin{itemize}
\item[(i)] If $q_{n}(\btheta)=\pi(\btheta)$, then $p_{{\rm acc},q_{n}}=\Theta_{p}(\varepsilon_{n}^{d}a_{n,\varepsilon}^{d-p})$
and $\Sigma_{{\rm IS},n}=\Theta_{p}(a_{n,\varepsilon}^{-2})$. 
\item[(ii)] If $q_{n}(\btheta)=\pi_{\rm ABC}(\btheta\mid \bs_{\rm obs},\varepsilon_{n})$, then
$p_{{\rm acc},q_{n}}=\Theta_{p}(\varepsilon_{n}^{d}a_{n,\varepsilon}^{d})$
and $\Sigma_{{\rm IS},n}=\Theta_{p}(a_{n,\varepsilon}^{p})$. 
\end{itemize}
In both cases $\hbh$ is asymptotically inefficient. \end{theorem} 
The result in part (ii) shows a difference from standard importance
sampling settings, where using the target distribution as the proposal
leads to an estimator with no Monte Carlo error.

The estimator $\hbh$ is asymptotically inefficient because the
Monte Carlo variance decays more slowly than $1/a_{n}^{2}$ as $n\rightarrow\infty$.
However this  is caused by different
factors in each case.

To see this, consider the acceptance probability of a value of $\theta$
and corresponding summary $\bs_{n}$ simulated in one iteration of
Algorithm \ref{alg:ISABC}. This acceptance probability depends on 
\begin{equation}
\frac{\bs_{n}-\bs_{\rm obs}}{\varepsilon_{n}}=\frac{1}{\varepsilon_{n}}\left[\{\bs_{n}-\bs(\btheta)\}+\{\bs(\btheta)-\bs(\btheta_{0})\}+\{\bs(\btheta_{0})-\bs_{\rm obs}\}\right],\label{eq:accp}
\end{equation}
where $\bs(\btheta)$, defined in Condition \ref{sum_conv}, is the limiting value of
$\bs_{n}$ as $n\rightarrow\infty$ if data is sampled from the model
for parameter value $\theta$. By Condition \ref{sum_conv}, the first and third bracketed
terms within the square brackets on the right-hand side are $O_{p}(a_{n}^{-1})$.
If we sample $\theta$ from the prior the middle term is $O_{p}(1)$,
and thus (\ref{eq:accp}) will blow up as $\varepsilon_{n}$ goes
to zero. Hence $p_{{\rm acc},\pi}$ goes to zero as $\varepsilon_{n}$ goes
to zero, which causes the estimate to be inefficient. If we sample
from the posterior, then by Theorem \ref{thm:1} we expect the middle
term to also be $O_{p}(a_{n}^{-1})$. Hence (\ref{eq:accp}) is well
behaved as $n\rightarrow\infty$, and  $p_{{\rm acc},\pi}$
is bounded away from zero, provided either $\varepsilon_{n}=\Theta(a_{n}^{-1})$
or $\varepsilon_{n}=\Omega(a_{n}^{-1})$.
%; though if $\varepsilon_{n}=o(a_{n}^{-1})$
%\eqref{eq:accp} will blow up as $n$ increases,
%which suggests that the acceptance probability will go to zero as $n\rightarrow\infty$.

However, if we use $\pi_{\rm ABC}(\btheta\mid \bs_{\rm obs},\varepsilon_{n})$ as
a proposal distribution, the estimates are still inefficient
due to an increasing variance of the importance weights: as $n$ increases
the proposal distribution is more and more concentrated around $\btheta_{0}$,
while $\pi$ does not change. 
%Therefore the weight, which is the ratio
%of $\pi_{\rm ABC}$ and $q_{\rm ABC}$, is increasingly skewed and causes
%$\Sigma_{{\rm IS},n}$ to go to $\infty$.

%\begin{rem}[Inefficiency of the negligible $\varepsilon_n$] \label{neglegible_bandwidth}
%As discussed after Theorem \ref{thm:1}, when $\varepsilon_{n}=o(a_{n}^{-1})$,
%the effect of the bias of approximate Bayesian computation is negligible. However, for any Monte Carlo
%algorithm making acceptance/rejection through $K(\bv)$, the acceptance
%probability with this choice of $\varepsilon_{n}$ goes to $0$ as
%$n\rightarrow\infty$. This can be seen from \eqref{eq:accp}. By Condition \ref{sum_conv} $\bs_{n}-\bs_{\rm obs}$ is $O_{p}(a_{n}^{-1})$
%and hence if $\varepsilon_{n}=o(a_n^{-1})$, \eqref{eq:accp}
%will blow-up, making the acceptance probability degenerate. In such
%a case, $N$ needs to increase with $n$ to compensate the decreasing
%acceptance rate.

%A corollary of this is that if we choose the bandwidth indirectly, by requiring a pre-determined proportion of acceptances, then the resulting $\varepsilon_n$ will not be in the 'negligible' regime.%\end{rem}

\subsection{Efficient Proposal Distributions}

\label{sec:proposal}

%Whilst using the prior or the posterior as a proposal distribution leads to asymptotically inefficient
%estimators, it will be seen that there exist practical choices for proposal distributions
%that avoid this inefficiency. 
Consider proposing the parameter value
from a location-scale family. That is our proposal is of the form
$\sigma_{n}\Sigma^{1/2}\bX+\mu_{n}$, where $\bX\sim q(\cdot)$, $E(\bX)=0$
and $\mbox{var}(\bX)=I_{p}$. This defines a general form of proposal density,
where the center, $\mu_{n}$, the scale rate, $\sigma_{n}$, the scale
matrix, $\Sigma$ and the base density, $q(\cdot)$, all need to be
specified. We will give conditions under which such a proposal density
results in estimators that are efficient.

%\empp{I found the description of the following condition hard to follow, so suggest a simpler presentation.}
Our results are based on an expansion of $\pi_{\rm ABC}(\btheta\mid \bs_{\rm obs},\varepsilon_{n})$.
Consider the rescaled random variables $\bt=a_{n,\varepsilon}(\btheta-\btheta_{0})$
and $\bv=\varepsilon_{n}^{-1}(\bs-\bs_{\rm obs})$. Recall that $\bT_{\rm obs}=a_{n}A(\btheta_{0})^{-1/2}\{\bs_{\rm obs}-\bs(\btheta_{0})\}$.
Define an unnormalised joint density of $\bt$ and $\bv$ as
\[
g_{n}(\bt,\bv;\tau)=\begin{cases}
\begin{array}{c}
N\Big[\{D\bs(\btheta_{0})+\tau\}\bt;a_{n}\varepsilon_{n}\bv+A(\btheta_{0})^{1/2}\bT_{\rm obs},A(\btheta_{0})\Big]K(\bv),\ \mbox{ }a_{n}\varepsilon_{n}\rightarrow c<\infty,\\
N\Big[\{D\bs(\btheta_{0})+\tau\}\bt;\bv+\frac{1}{a_{n}\varepsilon_{n}}A(\btheta_{0})^{1/2}\bT_{\rm obs},\frac{1}{a_{n}^{2}\varepsilon_{n}^{2}}A(\btheta_{0})\Big]K(\bv),\ \mbox{ }a_{n}\varepsilon_{n}\rightarrow\infty,
\end{array}\end{cases}
\]
and further define $g_{n}(\bt;\tau)=\int g_{n}(\bt,\bv;\tau)\,d\bv$. For large $n$, and for the rescaled variable $\bt$,
the leading term of $\pi_{\rm ABC}$ is then proportional to $g_{n}(\bt;0)$.
For both limits of $ a_{n}\varepsilon_{n}$, $g_{n}(\bt;\tau)$ is a continuous mixture of normal densities with the
kernel density determining the mixture weights. %Details of the expansion can be seen in the proof of Lemma 6 in the supplementary materials.

Our main theorem requires conditions on the proposal density. First, that $\sigma_{n}=a_{n,\varepsilon}^{-1}$ and that $c_{\mu}=\sigma_{n}^{-1}(\bmu_{n}-\btheta_{0})$ is $O_{p}(1)$. 
This ensures that under the scaling of $t$, as $n\rightarrow\infty$, the proposal is not increasingly over-dispersed compared to the target density, and the acceptance probability can be bounded away from zero. 
Second, that the proposal distribution is sufficiently heavy-tailed: 
\begin{condition} \label{approp_proposal}
There exist positive constants $m_{1}$ and $m_{2}$
satisfying $m_{1}^{2}I_{p}<Ds(\theta_{0})^{T}Ds(\theta_{0})$ and
$m_{2}I_{d}<A(\theta_{0})$, $\alpha\in(0,1)$, $\gamma\in(0,1)$ and $c\in(0,\infty)$,
such that for any $\lambda>0$,
\[
\sup_{t\in\mathbb{R}^{p}}\frac{N(t;0,m_{1}^{-2}m_{2}^{-2}\gamma^{-1})}{q\{\Sigma^{-1/2}(t-c)\}}<\infty,\ \sup_{t\in\mathbb{R}^{p}}\frac{\overline{K}^{\alpha}(\|\lambda t\|^2)}{q\{\Sigma^{-1/2}(t-c)\}}<\infty,\ 
\sup_{t\in\mathbb{R}^{p}}\frac{\overline{r}_{\rm max}(\|m_{1}m_{2}{\gamma}^{1/2}t\|^2)}{q\{\Sigma^{-1/2}(t-c)\}}<\infty,
\]
where $\overline{r}_{\rm max}(\cdot)$ satisfies $r_{\rm max}(v)=\overline{r}_{\rm max}(\|v\|_{\Lambda}^2)$, and for any random series $c_{n}$ in $\mathbb{R}^{p}$ satisfying
$c_{n}=O_{p}(1)$,
\[
\sup_{t\in\mathbb{R}^{p}}\frac{q(t)}{q(t+c_{n})}=O_{p}(1).
\]
\end{condition}

If we choose $\varepsilon_{n}=\Theta(a_{n}^{-1})$, the Monte
Carlo importance sampling variance for the accepted parameter values is $\Theta(a_{n}^{-2})$,
and has the same order as the variance of summary-based maximum likelihood estimator.%, giving an ABC estimator with non-degenerating asymptotic efficiency.
\begin{theorem}
\label{thm:3} Assume the conditions of Theorem \ref{thm:1}. If the proposal density $q_{n}(\btheta)$ is 
\[
\beta\pi(\btheta)+(1-\beta)\frac{1}{\sigma_{n}^{p}|\Sigma|^{1/2}}q\{\sigma_{n}^{-1}\Sigma^{-1/2}(\btheta-\bmu_{n})\},
\]
where $\beta\in(0,1)$, $q(\cdot)$ and $\Sigma$ satisfy Condition \ref{approp_proposal},
$\sigma_{n}=a_{n,\varepsilon}^{-1}$ and $c_{\mu}$ is $O_{p}(1)$, then
 $p_{{\rm acc},q_{n}}=\Theta_{p}(\varepsilon_{n}^{d}a_{n,\varepsilon}^{d})$
and $\Sigma_{{\rm IS},n}=O_{p}(a_{n,\varepsilon}^{-2})$. Then if $\varepsilon_{n}=\Theta(a_{n}^{-1})$,
$\textsc{AE}_{\hbh}=\Theta_{p}(1)$.

Furthermore, if $d=p$, $\textsc{AE}_{\hbh}=1-K/(N+K)$ for some constant
$K$. \end{theorem} 

The mixture with $\pi(\btheta)$ here is to control
the importance weight in the tail area \cite[]{Hesterberg:1995}. It is not clear whether
this is needed in practice, or is just a consequence of the approach
taken in the proof. %In practice, regarding the concern of the skewed importance weights, this implementation can be used to set a maximum value for the weights. But its safeguard effect may not be obvious since parameter values sampled in the tails are rarely accepted with a reasonable $\varepsilon$. This is illustrated in the numerical example in Section \ref{SV_example}.

Theorem \ref{thm:3} shows that with a good proposal distribution,
if the acceptance probability is bounded away from zero as $n$ increases,
the threshold $\varepsilon_{n}$ will have the preferred rate $\Theta(a_{n}^{-1})$.
This supports using the acceptance rate
to choose the threshold based on aiming for an appropriate proportion
of acceptances \cite[]{DelMoral:2012,biau2015new}.

In practice, $\sigma_{n}$ and $\mu_{n}$ need to be adaptive
to the observations since they depend on $n$. For $q(\cdot)$ and
$\Sigma$, the following proposition gives a practical suggestion
that satisfies Condition \ref{approp_proposal}. Let $T(\cdot;\gamma)$ be the multivariate
$t$ density with degree of freedom $\gamma$. The following result says that it is theoretically valid to choose any $\Sigma$ if a $t$ distribution is chosen as the base density.
\begin{proposition} \label{prop:1} Condition \ref{approp_proposal} is satisfied for
$q(\theta)=T(\btheta;\gamma)$ with any $\gamma>0$ and any $\Sigma$. 
\end{proposition} 
\begin{proof}
The first part of Condition \ref{approp_proposal} follows as the $t$-density is heavy tailed relative to the normal density, $\overline{K}(\cdot)$ and $\overline{r}_{\rm max}(\cdot)$. 
%from Conditions \ref{kernel_prop} and \ref{sum_approx_tail}
The second part can be verified easily.
\end{proof}

\subsection{Iterative Importance Sampling}

Taken together, Theorem \ref{thm:3} and Proposition \ref{prop:1}
suggest proposing from the mixture of $\pi(\btheta)$ and a $t$ distribution
with the scale matrix and center approximating those of $\pi_{\rm ABC}(\btheta)$.
We suggest the following iterative procedure, similar in spirit to that of \cite{beaumont2009adaptive}.

\begin{algo}%[!h]
{Iterative importance sampling approximate Bayesian computation} \label{alg:IIS} 

Input a mixture weight $\beta$, a sequence
of acceptance rates $\{p_{k}\}$, and a location-scale family. Set $q_1(\btheta)=\pi(\btheta)$.

For $k=1,\ldots,K$:
%  \vspace*{-20pt}
\begin{tabbing}
   \quad \quad 1. Run Algorithm 1 with simulation size $N_{0}$, proposal density $\beta\pi(\btheta)+(1-\beta)q_{k}(\btheta)$ and\\ 
   \quad \quad \quad acceptance rate $p_{k}$, and record the bandwidth $\varepsilon_{k}$.\\
   \quad \quad 2. If $\varepsilon_{k-1}-\varepsilon_{k}$ is smaller than some positive threshold, stop. Otherwise, let $\mu_{k+1}$ and $\Sigma_{k+1}$ \\
   \quad \quad \quad be the empirical mean and variance matrix of the weighted sample from step 1, and let\\ 
   \quad \quad \quad $q_{k+1}(\btheta)$ be the density with centre $\mu_{k+1}$ and variance matrix $2\Sigma_{k+1}$.\\
   \quad \quad 3. If $q_{k}(\btheta)$ is close to $q_{k+1}(\btheta)$ or $K=K_{\rm max}$, stop. Otherwise,
return to step $1$.
\end{tabbing}
After the iteration stops at the $K$\tsub{th} step, run Algorithm 1
with the proposal density $\beta\pi(\btheta)+(1-\beta)q_{K+1}(\btheta)$,
$N-KN_{0}$ simulations and $p_{K+1}$.
\end{algo}

In this algorithm, $N$ is the number of simulations allowed by the
computing budget, $N_{0}<N$ and $\{p_{k}\}$ is a sequence of acceptance rates, which we use to choose the bandwidth. 
The maximum value $K_{\rm max}$ of $K$ is set such that $K_{\rm max}N_{0}=N/2$. The rule for choosing
the new proposal distribution is based on approximating the mean and
variance of the density proportional to $\pi(\theta)f_{\rm ABC}(\bs_{\rm obs}\mid \theta,\varepsilon)^{1/2}$,
which is optimal \citep{fearnhead2012constructing}. It can be shown that
these two moments are approximately equal to the mean and twice the
variance of $\pi_{\rm ABC}(\btheta)$ respectively. For the mixture weight, $\beta$, we suggest a small value, and use $0.05$ in the simulation study below. 
%Since Algorithm \ref{alg:IIS} has the same simulation size as the rejection ABC and the additional calculations have negligible computational cost, the iterative procedure does not introduce additional cost. 

\section{Numerical Examples}

\subsection{Gaussian Likelihood with Sample Quantiles}

\label{Guassian_example} This examples illustrates the results in
Section 3 with an analytically tractable problem. Assume the observations
$\bY_{\rm obs}=(y_{1},\ldots,y_{n})$ follow the univariate normal distribution
$N(\mu,\sigma)$ with true parameter values $(1,2^{1/2})$. Consider
estimating the unknown parameter $(\mu,\sigma)$ with the uniform
prior in the region $[-10,10]\times[0,10]$ using Algorithm 1. The
summary statistic  is $(e^{\hq_{\alpha_{1}}/2},\ldots,e^{\hq_{\alpha_{d}}/2})$
where $\hq_{\alpha}$ is the sample quantile of $\bY_{\rm obs}$ for probability
$\alpha$.  
%Since the likelihood function and asymptotic distribution
%of the summary statistic are analytically available \citep{reiss2012approximate},
%the theoretical results in Theorem \ref{thm:1} and Proposition \ref{prop:dim_reduce}
%may be verified. 
%This summary statistic is illuminating because it
%is easy to change the information contained by changing the number
%of quantiles and it avoids the trivial case that $s(\theta)$ is a
%linear function of $\theta$.

\begin{figure}
\centering
\includegraphics[scale=0.9]{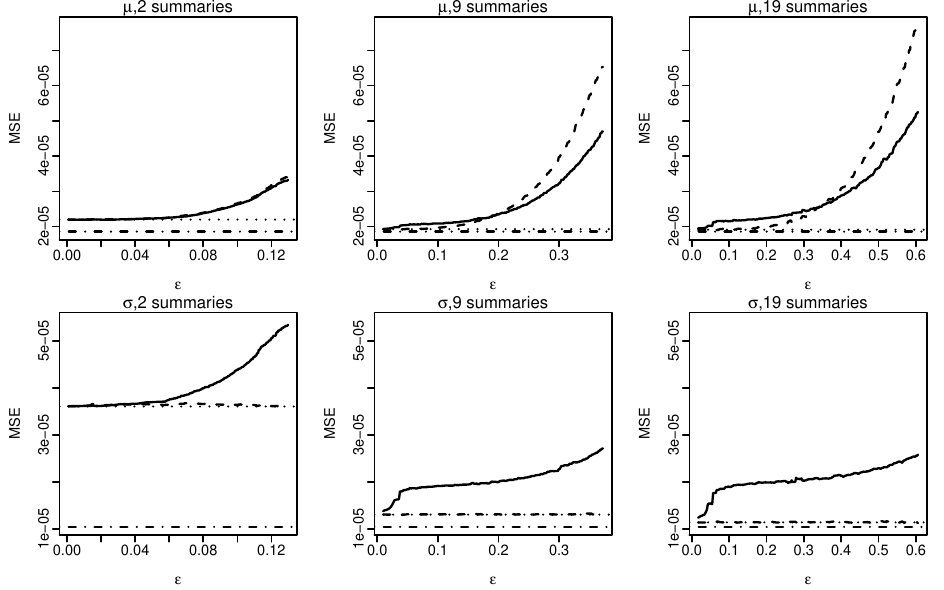} 
\caption{\label{ABC_for_Guassian}Illustration of results in Section 3. Mean square errors of point estimates for $200$ data sets are reported. Point estimates compared include $\theta_{\rm ABC}$ using the original summary statistic (solid) and the transformed summary statistic (dashed), the dimension of which is reduced to $2$ according to Proposition \ref{prop:dim_reduce}, the maximum likelihood estimates based on the original summary statistic (dotted) and the full data set (dash-dotted).}
\end{figure}

The results for data size $n=10^{5}$ are presented. Smaller sizes
from $10^{2}$ to $10^{4}$ show similar
patterns. The probabilities $\alpha_{1},\ldots,\alpha_{d}$ for calculating
quantiles are selected with equal intervals in $(0,1)$, and $d=2,9$
and $19$ were tested. 
In order to investigate the Monte Carlo error-free
performance, $N$ is chosen to be large enough that the Monte Carlo errors were negligible. 
We compare the performances of the ABC estimator
$\htheta$, the maximum likelihood estimator based on the summary statistics and the maximum likelihood estimator based on the full dataset. Since the dimension reduction
matrix $C$ in Proposition \ref{prop:dim_reduce} can be obtained
analytically, the performance of $\htheta$ using the original
$d$-dimension summary is compared with that using the $2$-dimension
summary. %, along with the same proposal distributions and acceptance rates since the values of $\varepsilon$ are not comparable. Since the values of $\varepsilon$ obtained using a single proposal distribution are limited, Various proposal distributions are used to obtain a wide enough interval. 
The results of mean square error are presented in Figure \ref{ABC_for_Guassian}.

The phenomena implied by Theorem \ref{thm:1} and Proposition \ref{prop:dim_reduce}
can be seen in this example, together with the limitations of these
results. First, ${E}\{\bh(\btheta)\mid \bs_{\rm obs}\}$, equivalent to
$\htheta$ with small enough $\varepsilon$, and the maximum likelihood estimator based on the same summaries, have similar accuracy. 
Second, when $\varepsilon$ is small, the mean square error
of $\htheta$ equals that of the maximum likelihood estimator based on the summary. When $\varepsilon$
becomes larger, for $d>2$ the mean square error increases more quickly than for $d=2$. This corresponds to the %`well-behaved' $\varepsilon$ case, and 
impact of the additional bias when $d>p$.
%As $\varepsilon$ increases further, the mean square error increases for all values of $d$.

For all cases, the two-dimensional summary obtained by projecting the original $d$ summaries is, for small $\varepsilon$, as accurate
as the maximum likelihood estimator given the original $d$ summaries.
This  indicates that the lower-dimensional summary contains the same information as the original one. 
%it does not mean that the transformation matrix $C$, even if analytically avaialble, can be used for reducing the dimension without other adjustment. 
 For larger $\varepsilon$, the performance of the reduced-dimension
summaries is not stable, and is in fact worse than the original
summaries for estimating $\mu$.
This deterioration is caused by the bias of $\htheta$, which for larger $\varepsilon$, is dominated
by higher order terms in $\varepsilon$ which could be ignored in our asymptotic results.
%This is due to the second order behaviour of $\theta_{\rm ABC}$, which
%becomes important for larger $\varepsilon$. %, is changed by multiplying $C$ to the summary statistics. 
%This suggests using other techniques for reducing the bias, e.g. the
%regression adjustment, together with the dimension-reduction matrix
%for more stable behaviour.

\subsection{Stochastic Volatility with AR(1) Dynamics}

\label{SV_example}
We consider a stochastic volatility model from \cite{sandmann1998estimation} for the de-meaned returns of a portfolio. Denote this return for the 
$t$th time-period as $y_t$. Then
\[
%\begin{cases}
x_{t}  =\phi x_{t-1}+\eta_{t},\ \eta_{t}\sim N(0,\sigma_{\eta}^{2}); ~~
%\]
%\[
y_{t}  =\overline{\sigma}e^{{x_{t}}/{2}}\xi_{n},\ \xi_{t}\sim N(0,1),
%\end{cases}
\]
where $\eta_{t}$ and $\xi_{t}$ are independent, and $x_t$ is a latent state that quantifies
the level of volatility for time-period $t$.
%, $y_{n}$ is the
%demeaned return of a portfolio obtained by subtracting the average
%of all returns from the actual return and $\overline{\sigma}$ is
%the average volatility level. 
By the transformation $y_{t}^{*}=\log y_{t}^{2}$
and $\xi_{t}^{*}=\log\xi_{t}^{2}$, the observation equation in the 
state-space model can be transformed to 
\begin{equation}
%\begin{cases}
%x_{n} & =\phi x_{n-1}+\eta_{n},\ \eta_{n}\sim N(0,\sigma_{\eta}^{2})\\
y_{n}^{*}  =2\log\overline{\sigma}+x_{n}+\xi_{n}^{*},\ \exp(\xi_{n}^{*})\sim\chi_{1}^{2},
%\end{cases}\label{SV_model}
\end{equation}
which is linear and non-Gaussian.

\begin{figure}[t]
\begin{centering}
\includegraphics[scale=0.8]{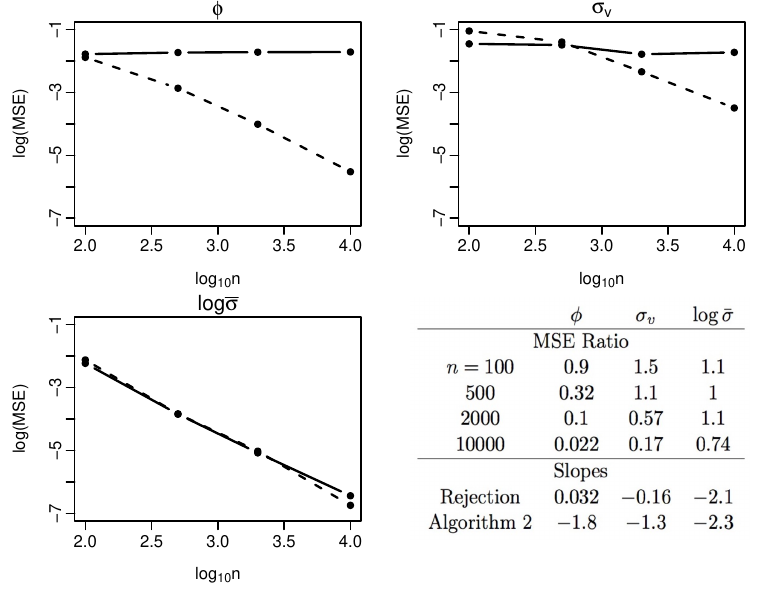} 
\par\end{centering}

 \caption{Comparisons of rejection (solid) and iterative importance sampling (dashed) versions of approximate Bayesian computation.
For each $n$,
the logarithm of the average mean square error across $100$ datasets is reported. 
For each dataset, the Monte Carlo sample size is $40000$. Ratios of mean square errors of the two methods
are given in the table, and smaller values indicate better performance
of iterative importance sampling. For each polyline in the plots, a line is fitted and the slope is reported in the table. Smaller values indicate faster decrease of the mean square error.}
%\vspace{-10pt}
 \label{ABC_for_SV} 
\end{figure}
Approximate Bayesian computation can be used to obtain an off-line estimator for the
unknown parameters of this model.
%, which is recently discussed
%by \cite{martin2014approx}. 
Here we illustrate the effectiveness
of iteratively choosing the importance proposal for large $n$ by
comparing with rejection sampling. % the rejection and the iterative
%importance sampling versions. 
In the iterative algorithm, a $t$ distribution with 5 degrees
of freedom is used to construct $q_{k}$.

Consider estimating the parameter $(\phi,\sigma_{\eta},\log\overline{\sigma})$
under a uniform prior in the region $[0,1)\times[0.1,3]\times[-10,-1]$.
The setting with the true parameter $(\phi,\sigma_{\eta},\log\overline{\sigma})=(0.9,0.675,-4.1)$
is studied. %, which is motivated by the empirical studies. 
We use a three-dimensional summary statistic that stores the mean, variance and lag-one autocovariance of the transformed data. 
If
there were no noise in the state equation for $\xi_{n}^{*}$, % in \eqref{SV_model},
then this would be a sufficient statistic of $\bY^{*}$,
and hence is a natural choice for the summary statistic. The uniform
kernel is used in the accept-reject step.

We evaluate rejection sampling and iterative importance sampling methods on data of length $n=100,500,2000$ and $10000$; and use $N=40000$ Monte Carlo simulations. 
For iterative importance sampling, the sequence $\{p_{k}\}$
has the first five values decreasing linearly from $5\%$ to $1\%$,
and later values being $1\%$. We further set $N_0=2000$, and $K_{\rm max}=10$. For the rejection sampler acceptance probabilities of both $5\%$ and $1\%$
were tried and $5\%$ was chosen as it gave better performance.
The simulation results are shown
in Figure \ref{ABC_for_SV}. 

For all parameters, iterative importance sampling shows increasing
advantage over rejection sampling as $n$ increases. For larger $n$, the iterative
procedure obtains a center for proposals closer to the true parameter
and a bandwidth that is smaller than those used for rejection sampling.
%, and the comparison
%becomes more significant when $n$ increases. 
These contribute to the more accurate estimators. 
%For smaller $n$, both perform similarly,
%since when the summary statistic is not accurate enough, the ABC posterior
%is not much different from the prior, and the benefit of sampling
%from a slightly better proposal does not compensate the increased
%Monte Carlo variance from the importance weight. 
It is
easy to estimate $\log\bar{\sigma}$, since the expected summary statistic
$\widetilde{E}(\bY^{*})$ is roughly linear in $\log\bar{\sigma}$.
Thus iterative importance sampling has less of an advantage over rejection sampling when
estimating this parameter. 
%Finally, the performance both
%with and without the mixture for the proposal density are similar.%, showing that the skewness of the importance weight does not have significant effect. 

\section{Discussion}

Our results suggest one can obtain efficient estimates using Approximate Bayesian Computation with a fixed Monte Carlo sample size
as $n$ increases. Thus the computational complexity of approximate Bayesian computation will just be the complexity of simulating
a sample of size $n$ from the underlying model.

Our results on the Monte Carlo accuracy of approximate Bayesian computation considered the importance sampling implementation given in Algorithm \ref{alg:ISABC}. If we do not use the uniform kernel,
then there is a simple improvement on this algorithm, that absorbs the accept-reject probability within the importance sampling weight. A simple Rao--Blackwellisation argument then shows that this
leads to a reduction in Monte Carlo variance, so our positive results about the scaling of approximate Bayesian computation with $n$ will also immediately apply to this implementation. 

Similar positive Monte Carlo results are likely to apply to Markov chain Monte Carlo implementations of approximate Bayesian computation.
A Markov chain Monte Carlo version will be efficient provided the acceptance probability
does not degenerate to zero as $n$ increases. However at stationarity,
it will propose parameter values from a distribution close to
the approximate Bayesian computation posterior density, and Theorems \ref{thm:2} and \ref{thm:3} suggest that for
such a proposal distribution the acceptance probability will
be bounded away from zero.

Whilst our theoretical results suggest that point estimates based
on approximate Bayesian computation have good properties, they do not suggest that
the approximate Bayesian computation posterior is a good approximation to the true posterior. In fact,
\cite{Frazier:2017} show it will over-estimate uncertainty if $\varepsilon_n=O(a_n^{-1})$. 
However, \cite{Li/Fearnhead:2017}
show that using regression methods \cite[]{Beaumont:2002} to post-process approximate Bayesian computation output can lead to both efficient point estimation and accurate quantification of uncertainty.

\section*{Acknowledgment}

This work was support by the Engineering and Physical Sciences Research Council.%, grant EP/K014463.

\section*{Supplementary Material}

\section{Proof of Results from Section 3} \label{App:A}

\subsection{Overview and Notation} \label{S:overview}

We first give an overview of the proof to Theorem \ref{thm:1}. The convergence of the maximum likelihood estimator based on the summary follows almost immediately from \cite{creel2013indirect}. The minor extensions
we used are summarized in Lemmas \ref{lem2:theta_MLE} and \ref{lem4:h_converg} below. 

The main challenge with Theorem \ref{thm:1} are the results about the posterior mean of approximate Bayesian computation. 
For the convergence of posterior means of approximate Bayesian computation we need to consider convergence of integrals over the parameter space, $\mathbbm{R}^{p}$. 
We will divide $\mathbbm{R}^{p}$ into $B_{\delta}=\{\theta:\|\theta-\theta_{0}\|<\delta\}$
and $B_{\delta}^{c}$ for some $\delta<\delta_{0}$, and introduce the notation 
$\pi(h)=\int h(\theta)\pi(\theta)f_{\rm ABC}(s_{{\rm obs}}\mid\theta)\,d\theta$.
The posterior mean of approximate Bayesian computation is $h_{\rm ABC}=\pi(h)/\pi(1)$. 
We can write $\pi(h)$, say, as $\pi(h)= \pi_{B_{\delta}}(h)+\pi_{B_{\delta}^{c}}(h)$, where
\[
\pi_{B_{\delta}}(h)=\int_{B_{\delta}}h(\theta)\pi(\theta)f_{\rm ABC}(s_{{\rm obs}}\mid\theta)\,d\theta,\quad \pi_{B_{\delta}^{c}}(h)=\int_{B_{\delta}^{c}}h(\theta)\pi(\theta)f_{\rm ABC}(s_{{\rm obs}}\mid\theta)\,d\theta.
\]

As $n\rightarrow \infty$ the posterior distribution of approximate Bayesian computation concentrates around $\theta_0$. The first step of our proof is to show that, as a result, the contribution 
that comes from integrating over $B_{\delta}^{c}$ can be ignored. Hence we need consider only $\pi_{B_{\delta}}(h)/\pi_{B_{\delta}}(1)$.

Second, we perform a Taylor expansion of $h(\theta)$ around $\theta_0$. Let $Dh(\theta)$ and $Hh(\theta)$ denote the vector of first derivatives and the matrix of second derivatives of $h(\theta)$ respectively. Then
\[
 h(\theta)=h(\theta_0)+Dh(\theta_0)^T(\theta-\theta_0)+\frac{1}{2}(\theta-\theta_0)^THh(\theta_*)(\theta-\theta_0),
\]
for some $\theta_*$, that depends on $\theta$ and that satisfies $||\theta_*-\theta_0||<||\theta-\theta_0||$. We plug this into $\pi_{B_{\delta}}(h)$, but re-express the integrals in term of the 
rescaled random vector
\[
 t(\theta)=a_{n,\varepsilon}(\theta-\theta_{0}),
\]
and let $t(B_{\delta})$ be the set $\{\phi:\phi=t(\theta)\text{ for some }\theta\in B_{\delta}\}$. This gives
\begin{align}
\frac{\pi_{B_{\delta}}(h)}{\pi_{B_{\delta}}(1)}=h(\theta_{0})+a_{n,\varepsilon}^{-1}Dh(\theta_{0})^{T}\frac{\pi_{B_{\delta}}(t)}{\pi_{B_{\delta}}(1)}+\frac{1}{2}a_{n,\varepsilon}^{-2}
\frac{\pi_{B_{\delta}}\{t^{T}Hh(\theta_{t})t\}}{\pi_{B_{\delta}}(1)},\label{h_expand}
\end{align}
where we write $t$ for $t(\theta)$, and  $\theta_{t}$ is the value $\theta_*$ from remainder term in the Taylor expansion for $h(\theta)$. We use the notation $\theta_t$ to emphasize its
dependence on $t$, and note that $\theta_t$ belongs to $B_{\delta}$. 

Let $\ftil_{\rm ABC}(s_{{\rm obs}}\mid\theta)=\int\ftil_{n}(s_{{\rm obs}}+\varepsilon_{n}v\mid\theta)K(v)\,dv$, which is the likelihood approximation  that we get if
we replace the true likelihood by its Gaussian limit, and define $\tpi_{B_{\delta}}(h)=\int_{B_{\delta}}h(\theta)\pi(\theta)\ftil_{\rm ABC}(s_{{\rm obs}}\mid\theta)\,d\theta$. Our third step is to re-write
(\ref{h_expand}) as
\begin{eqnarray*}
 \frac{\pi_{B_{\delta}}(h)}{\pi_{B_{\delta}}(1)}&=&h(\theta_{0})+a_{n,\varepsilon}^{-1}Dh(\theta_{0})^{T}\frac{\tpi_{B_{\delta}}(t)}{\tpi_{B_{\delta}}(1)} 
  + a_{n,\varepsilon}^{-1}Dh(\theta_{0})^{T}\left\{ \frac{\tpi_{B_{\delta}}(t)}{\tpi_{B_{\delta}}(1)}-\frac{\pi_{B_{\delta}}(t)}{\pi_{B_{\delta}}(1)} 
 \right\} \\ & & + \frac{1}{2}a_{n,\varepsilon}^{-2} \frac{\pi_{B_{\delta}}\{t^{T}Hh(\theta_{t})t\}}{\pi_{B_{\delta}}(1)}.
\end{eqnarray*}
We bound the size of the last two terms, so that asymptotically $h_{\rm ABC}$ behaves as
\[
 h(\theta_{0})+a_{n,\varepsilon}^{-1}Dh(\theta_{0})^{T}\frac{\tpi_{B_{\delta}}(t)}{\tpi_{B_{\delta}}(1)}.
\]
If we introduce the density $g_n(t,v)$, defined as $g_n(t,v,\tau)$ in Section \ref{sec:proposal} of the main text but with $\tau=0$, so
 \[
g_{n}(t,v)\propto\begin{cases}
\begin{array}{c}
N\Big\{Ds(\theta_{0})t;a_{n}\varepsilon_{n}v+A(\theta_{0})^{1/2}T_{\rm obs},A(\theta_{0})\Big\}K(v),\quad a_{n}\varepsilon_{n}\rightarrow c<\infty,\\
N\Big\{Ds(\theta_{0})t;v+\frac{1}{a_{n}\varepsilon_{n}}A(\theta_{0})^{1/2}T_{\rm obs},\frac{1}{a_{n}^{2}\varepsilon_{n}^{2}}A(\theta_{0})\Big\}K(v),\quad a_{n}\varepsilon_{n}\rightarrow\infty,
\end{array}\end{cases}
\]
then we can show that
\[
\frac{\widetilde{\pi}_{B_{\delta}}(t)}{\widetilde{\pi}_{B_{\delta}}(1)}\approx \frac{\int_{t(B_{\delta})}
\int_{\mathbb{R}^{d}}tg_{n}(t,v)\,dtdv}{\int_{t(B_{\delta})}\int_{\mathbb{R}^{d}}g_{n}(t,v)\,dtdv},
\]
with a remainder that can be ignored. Putting this together, we get that asymptotically $h_{\rm ABC}$ is 
\[
 h(\theta_{0})+a_{n,\varepsilon}^{-1}Dh(\theta_{0})^{T}\frac{\int_{t(B_{\delta})}
\int_{\mathbb{R}^{d}}tg_{n}(t,v)\,dtdv}{\int_{t(B_{\delta})}\int_{\mathbb{R}^{d}}g_{n}(t,v)\,dtdv},
\]
and the proof finishes by calculating the form of this.

A recurring theme in the proofs for the bounds on the various remainders is the need to bound
expectations of polynomials of either the rescaled parameter $t$, or a rescaled difference in the summary statistic
from $s_{{\rm obs}}$, or both. Later we will present a lemma, stated in terms of a general polynomial, that is used repeatedly to obtain 
the bounds we need. 

To define this we need to introduce a set of suitable polynomials.  
For any integer $l$ and vector $x$, if a scalar function of $x$
has the expression $\sum_{i=0}^{l}\alpha_{i}(x,n)^{T}x^{i}$, where
for each $i$, $x^{i}$ denotes the vector with all monomials of $x$
with degree $i$ as elements and $\alpha_{i}(x,n)$ is a vector of
functions of $x$ and $n$, we denote it by $P_{l}(x)$. Let $\mathbb{P}_{l,x}$
be the set
\[
\{P_{l}(x):\text{for all }i\leq l,\ \text{as }n\rightarrow\infty,\ \alpha_{i}(x,n)=%O(1)\text{ or }
O_{p}(1)\text{ holds uniformly in }x\}%\text{ if it is random}\}.
\]
To simplify the notations, for two vectors $x_{1}$ and $x_{2}$,
$P_{l}\{(x_{1}^{T},x_{2}^{T})^{T}\}$ and $\mathbb{P}_{l,(x_{1}^{T},x_{2}^{T})^{T}}$
are written as $P_{l}(x_{1},x_{2})$ and $\mathbb{P}_{l,(x_{1},x_{2})}$. Where the specific form
of the polynomial does not matter, and we only use the fact that it lies in $\mathbb{P}_{l,x}$,
we will often simplify expressions by writing it as $P_l(x)$.

\subsection{Proof of Theorem \ref{thm:1}} \label{S:Thm1}

For the maximum likelihood estimator based on the summary, \cite{creel2013indirect} gives the central limit theorem
for $\hat{\theta}_{\mbox{\scriptsize \rm MLES}}$ when $a_{n}={n}^{1/2}$
and $\mathcal{P}$ is compact. According to the proof in \cite{creel2013indirect},
extending the result to the general $a_{n}$ is straightforward. Additionally,
we give the extension for general $\mathcal{P}$. 
\begin{lemma} \label{lem2:theta_MLE} Assume
Conditions \ref{par_true},\ref{sum_conv}-\ref{sum_approx_tail}.
Then $a_{n}(\hat{\theta}_{\mbox{\scriptsize \rm MLES}}-\theta_{0})\rightarrow N\{0,I^{-1}(\theta_{0})\}$ in distribution
as $n\rightarrow\infty$. \end{lemma} 
Given Condition \ref{h_Cdifferentiable},
by Lemma \ref{lem2:theta_MLE} and the delta method \cite[]{lehmann2004elements}, the convergence
of the maximum likelihood estimator for general $h(\theta)$ holds as follows. 
\begin{lemma}
\label{lem4:h_converg} Assume the conditions of Lemma \ref{lem2:theta_MLE}
and Condition \ref{h_Cdifferentiable}. Then $a_{n}\{h(\hat{\theta}_{\mbox{\scriptsize \rm MLES}})-h(\theta_{0})\}\rightarrow N\{0,Dh(\theta_{0})^{T}I^{-1}(\theta_{0})Dh(\theta_{0})\}$ in distribution as $n\rightarrow\infty$.
\end{lemma}

The following lemmas are used for the result about the posterior mean of approximate Bayesian computation, proofs of these are given in Section \ref{S:Proof_Lem}. Our first lemma is used to justify
ignoring integrals over ${B_{\delta}^{c}}$.
\begin{lemma}
\label{marglik_ignored} Assume Conditions \ref{kernel_prop}, \ref{h_Cdifferentiable}--\ref{sum_approx_tail}. Then
for any $\delta<\delta_0,$ $\pi_{B_{\delta}^{c}}(h)=O_{p}(e^{-a_{n,\varepsilon}^{\alpha_{\delta}}c_{\delta}})$
for some positive constants $c_{\delta}$ and $\alpha_{\delta}$ depending
on $\delta$. 
\end{lemma}

The following lemma is used to calculate the form of 
\[
 \frac{\int_{t(B_{\delta})}
\int_{\mathbb{R}^{d}}tg_{n}(t,v)\,dtdv}{\int_{t(B_{\delta})}\int_{\mathbb{R}^{d}}g_{n}(t,v)\,dtdv},
\]
which is the leading term for $\{h_{\rm ABC}-h(\theta_0)\}$.%, when $d>p$.
\begin{lemma} \label{t_mixnorm_val} 
%Consider notations from Lemma
%\ref{poly_mixnorm_bound} and assume $d>p$. If conditions of Lemma
%\ref{poly_mixnorm_bound} are satisfied, 
Assume Condition \ref{kernel_prop}. Let $c$ be a constant vector, $\{k_{n}\}$ be
a series converging to $k_{\infty}\in(0,\infty]$ and $\{b_{n}'\}$
be a series converging to a non-negative constant. Let $b_{n}=\mathbbm{1}_{\{k_{\infty}=\infty\}}+b_{n}'\mathbbm{1}_{\{k_{\infty}<\infty\}}$.
Then for any $d\times p$
constant matrix $A$ and any $d\times d$ constant matrix $B$, 
\[
\int_{\mathbb{R}^{p}}\int_{\mathbb{R}^{d}}t\frac{N(At;B_{n}v+\frac{1}{k_{n}}c,\frac{1}{k_{n}^{2}}I_{d})K(v)}{\int_{\mathbb{R}^{p}}\int_{\mathbb{R}^{d}}N(At;B_{n}v+\frac{1}{k_{n}}c,\frac{1}{k_{n}^{2}}I_{d})K(v)\,dtdv}\,dtdv=\frac{1}{k_{n}}\left\{ (A^{T}A)^{-1}A^{T}c+R(A,B_{n},k_{n},c)\right\} ,
\]
where $B_{n}=b_{n}B$, the expression of $R(c;A,B_{n},k_{n})$
is stated in the proof. Specifically, $R(A,B_{n},k_{n},c)=o(1)$ when $B_{n}=o(1)$ and $O(1)$ otherwise.
\end{lemma} 

Our final two lemmas are used to bound the remainder terms in the expansion for $h_{\rm ABC}$ we presented in Section \ref{S:overview}.
\begin{lemma} \label{posterior_normal}Assume Conditions \ref{par_true},
\ref{kernel_prop} and \ref{sum_conv}
hold. If $\varepsilon_{n}=o({a_{n}}^{-1/2})$, there exists a $\delta<\delta_{0}$
such that 
\begin{align}
 & \widetilde{\pi}_{B_{\delta}}(1)=a_{n,\varepsilon}^{d-p}\Big\{\pi(\theta_{0})\int_{t(B_{\delta})}\int_{\mathbb{R}^{d}}g_{n}(t,v)\,dvdt+O_{p}(a_{n,\varepsilon}^{-1})+O_{p}(a_{n}^{2}\varepsilon_{n}^{4})\Big\},\nonumber \\
 & \int_{t(B_{\delta})}\int_{\mathbb{R}^{d}}g_{n}(t,v)\,dtdv=\Theta_{p}(1),\nonumber \\
 & \frac{\widetilde{\pi}_{B_{\delta}}(t)}{\widetilde{\pi}_{B_{\delta}}(1)}=\frac{\int_{t(B_{\delta})}\int_{\mathbb{R}^{d}}tg_{n}(t,v)\,dtdv}{\int_{t(B_{\delta})}\int_{\mathbb{R}^{d}}g_{n}(t,v)\,dtdv}+O_{p}(a_{n,\varepsilon}^{-1})+O_{p}(a_{n}^{2}\varepsilon_{n}^{4}),\label{eq:post_lim2}
\end{align}
and $\widetilde{\pi}_{B_{\delta}}\{P_{2}(t)\}/\widetilde{\pi}_{B_{\delta}}(1)=O_{p}(1)$
for any $P_{2}(t)\in\mathbb{P}_{2,t}$. \end{lemma}

\begin{lemma}\label{lem:norm_remain} Assume the conditions of Lemma \ref{posterior_normal} and Conditions \ref{h_Cdifferentiable}
and \ref{sum_approx}. Then if $\varepsilon_{n}=o({a_{n}}^{-1/2})$,
there exists a $\delta<\delta_{0}$ such that
\begin{equation}
\frac{\pi_{B_{\delta}}(h)}{\pi_{B_{\delta}}(1)}=h(\theta_{0})+a_{n,\varepsilon}^{-1}Dh(\theta_{0})^{T}\left\{\frac{\tpi_{B_{\delta}}(t)}{\tpi_{B_{\delta}}(1)}+O_{p}(\alpha_{n}^{-1})\right\}+
\frac{1}{2}a_{n,\varepsilon}^{-2}\left[\frac{\tpi_{B_{\delta}}\{t^{T}Hh(\theta_{t})t\}}{\tpi_{B_{\delta}}(1)}+O_{p}(\alpha_{n}^{-1})\right],\label{norm_expand7}
\end{equation}
%\eqref{eq:norm_remain} holds.
\end{lemma}

Now we are ready to prove Theorem \ref{thm:1}.

\begin{proof}[of Theorem \ref{thm:1}]
The convergence of the maximum likelihood estimator based on the summary is given by Lemma \ref{lem2:theta_MLE}
and Lemma \ref{lem4:h_converg}. 

We now focus on the convergence for the posterior mean of approximate Bayesian computation. The convergence of the posterior mean given the summaries follows from
a similar, but simpler, argument and is omitted.

We can bound $t^TH(\theta_t)t$ for $\theta$ in $B_{\delta}$ by the quadratic $t^TH_{max}t$, where $H_{max}$ is an upper bound
on $H(\theta_t)$ for $\theta_t$ in $B_{\delta}$. This means that 
\[
 \tpi_{B_{\delta}}\{t^{T}Hh(\theta_{t})t\}=O(1).
\]
Together with Lemmas \ref{marglik_ignored}, \ref{posterior_normal} and \ref{lem:norm_remain}, 
we then have
the expansion
\[
h_{\rm ABC}=h(\theta_{0})+a_{n,\varepsilon}^{-1}Dh(\theta_{0})^{T}\left\{ 
\frac{\int_{t(B_{\delta})\times\mathbb{R}^{d}}tg_{n}(t,v)\,dtdv}{\int_{t(B_{\delta})\times\mathbb{R}^{d}}g_{n}(t,v)\,dtdv}+O_{p}(a_{n,\varepsilon}^{-1})+O_{p}(a_{n}^{2}\varepsilon_{n}^{4})+O_{p}(\alpha_{n}^{-1})
\right\} .
\]

The analytical form of the integral in the above expansion, which we will denote
by $E_{g_{n}}(t)$, can be obtained by applying Lemma \ref{t_mixnorm_val}
with $A=A(\theta_{0})^{-1/2}DS(\theta_{0})$, $c=T_{{\rm obs}}$,
\begin{align*}
 \ensuremath{B_{n}=\begin{cases}
a_{n}\varepsilon_{n}A(\theta_{0})^{-1/2}, & c_{\varepsilon}<\infty,\\
A(\theta_{0})^{-1/2}, & c_{\varepsilon}=\infty,
\end{cases}}\quad \ensuremath{k_{n}=\begin{cases}
1, & c_{\varepsilon}<\infty,\\
a_{n}\varepsilon_{n}, & c_{\varepsilon}=\infty.
\end{cases}}
\end{align*}
It can be seen that $E_{g_{n}}(t)$ is $\Theta_{p}(k_{n}^{-1})$,
and the remainder term, $O_{p}(a_{n,\varepsilon}^{-1})+O_{p}(a_{n}^{2}\varepsilon_{n}^{4})+O_{p}(\alpha_{n}^{-1})$,
is $o_{p}(1)$ as $\varepsilon_{n}=o(a_{n}^{-3/5})$
and $\alpha_{n}^{-1}=o(a_{n}^{-2/5})$. Then since $a_{n,\varepsilon}^{-1}k_{n}^{-1}=a_{n}^{-1}$,
we have 
\begin{align}
 & a_{n}\{h_{\rm ABC}-h(\theta_{0})\}\nonumber \\
= & Dh(\theta_{0})^{T}\Big[\Big\{ Ds(\theta_{0})^{T}A(\theta_{0})^{-1}Ds(\theta_{0})\Big\}^{-1}Ds(\theta_{0})^{T}A(\theta_{0})^{-1/2}T_{{\rm obs}}+R_{n}(a_{n}\varepsilon_{n},T_{\rm obs})\Big]+o_{p}(1), \label{eq:thm1_1}
\end{align}
where $R_{n}(a_{n}\varepsilon_{n},T_{\rm obs})$ is $Dh(\theta_{0})^{T}R(A,B_{n},k_{n},c)$
with $R(A,B_{n},k_{n},c)$ defined in Lemma \ref{t_mixnorm_val}.
We can interpret $R_{n}(a_{n}\varepsilon_{n},T_{\rm obs})$
as the extra variation brought by $\varepsilon_{n}$: $a_{n}[h_{\rm ABC}-{E}\{h(\theta)\mid s_{{\rm obs}}\}]$.

By the delta method, the first term in the right hand side of \eqref{eq:thm1_1}
converges to $I(\theta_{0})^{-1/2}Z$. For the second term, since
$A(A^{T}A)^{-1}A^{T}$ is a projection matrix, by eigen decompositition
\[
I-A(A^{T}A)^{-1}A^{T}=U\left(\begin{array}{cc}
0 & 0\\
0 & I_{d-p}
\end{array}\right)U^{T},\ (A^{T}A)^{-1/2}A^{T}=\left(\begin{array}{cc}
I_{p} & 0\\
0 & 0
\end{array}\right)U^{T},
\]
where $U$ is an orthogonal matrix. For a vector $x$, let $x_{k_{1}:k_{2}}$
be the $(k_{2}-k_{1}+1)$-dimension vector containing the $k_{1}$th\textendash $k_{2}$th
coordinates of $x$. Let $v'=U^{T}A(\theta_{0})^{-1/2}v$, and $T_{{\rm obs}}'=U^{T}T_{{\rm obs}}$.
Then $R_{n}(a_{n}\varepsilon_{n},T_{\rm obs})$ can be written as 
\begin{align}
 & R_{n}(a_{n}\varepsilon_{n},T_{\rm obs})\nonumber \\
= & Dh(\theta_{0})^{T}(A^{T}A)^{-1/2}a_{n}\varepsilon_{n}\frac{\int v_{1:p}'N\{v_{(p+1):d}';-\frac{1}{a_{n}\varepsilon_{n}}T_{{\rm obs},(p+1):d}',\frac{1}{a_{n}^{2}\varepsilon_{n}^{2}}I_{d-p}\}
K\{A(\theta_{0})^{1/2}Uv'\}\,dv'}
{\int N\{v_{(p+1):d}';-\frac{1}{a_{n}\varepsilon_{n}}T_{{\rm obs},(p+1):d}',\frac{1}{a_{n}^{2}\varepsilon_{n}^{2}}I_{d-p}\}K\{A(\theta_{0})^{1/2}Uv'\}\,dv'}.\label{eq:thm1_2}
\end{align}
Denote the weak limit of $R_{n}(a_{n}\varepsilon_{n},T_{\rm obs})$ as
$R(c_{\varepsilon},Z)$. When $d=p$, obviously $R_{n}(a_{n}\varepsilon_{n},T_{\rm obs})=0$
and therefore $R(c_{\varepsilon},Z)=0$. When $d>p$, if $\varepsilon_{n}=o(1/a_{n})$,
$R_{n}(a_{n}\varepsilon_{n},T_{\rm obs})=o_{p}(1)$ by Lemma \ref{t_mixnorm_val}
and therefore $R(c_{\varepsilon},Z)=0$. 
%When $A(\theta_{0})$ is
%diagonal, for fixed $v_{(p+1):d}'$, $K\{A(\theta_{0})^{1/2}Uv'\}$
%as a function of $v_{1:p}'$ i
When the covariance matrix of $K(\cdot)$ is $c^{2}A(\theta_{0})$,
for constant $c>0$,  $K(v)\propto\overline{K}\{c\|A(\theta_{0})^{-1/2}v\|^{2}\}$.
Then $K\{A(\theta_{0})^{1/2}Uv'\}$ in \eqref{eq:thm1_2} can be replaced
by $\overline{K}(c\|v'\|^{2})$ and for fixed $v_{(p+1):d}'$, the
integrand in the numerator, as a function of $v_{1:p}'$, is symmetric
around zero. Therefore $R_{n}(a_{n}\varepsilon_{n},T_{\rm obs})=0$ and
$R(c_{\varepsilon},Z)=0$.

Otherwise, $R_{n}(a_{n}\varepsilon_{n},z)$
is not necessarily zero. Since for any $n$, $R_{n}(a_{n}\varepsilon_{n},z)$
as a function of $z$ is symmetric around $0$, $R(c_{\varepsilon},z)$
is also symmetric and $R(c_{\varepsilon},Z)$ has mean zero. Since
$I^{-1}(\theta_{0})$ is the Cramer-Rao lower bound, $\mbox{var}\{I(\theta_{0})^{-1/2}Z+R(c_{\varepsilon},Z)\}\geq I^{-1}(\theta_{0})$.

For (i), the asymptotic normality holds for $h(\hat{{\theta}})$ by
Lemma \ref{lem4:h_converg}. \end{proof}

\subsection{Proof of Lemmas}\label{S:Proof_Lem}

Here we give the proofs of lemmas from Section \ref{S:Thm1}.

\begin{proof}[of Lemma \ref{marglik_ignored}] It is sufficient to show that for any $\delta$, $\sup_{\theta\in B_{\delta}^{c}}f_{\rm ABC}(s_{{\rm obs}}\mid\theta)=O_{p}(e^{-a_{n,\varepsilon}^{\alpha_{\delta}}c_{\delta}})$.
By dividing $\mathbb{R}^{d}$ into $\{v:\|\varepsilon_{n}v\|\leq\delta'/3\}$ and its complement, we have 
\begin{eqnarray*}
\lefteqn{\sup_{\theta\in B_{\delta}^{c}}f_{\rm ABC}(s_{{\rm obs}}\mid\theta)=  \sup_{\theta\in B_{\delta}^{c}}\int_{\mathbb{R}^{d}}f_{n}(s_{{\rm obs}}+\varepsilon_{n}v\mid\theta)K(v)\,dv }\\
&\leq & \sup_{\theta\in B_{\delta}^{c}\backslash\mathcal{P}_{0}^{c}}\left\{\sup_{\|s-s_{{\rm obs}}\|\leq\delta'/3}f_{n}(s\mid\theta)\right\}+\sup_{\theta\in\mathcal{P}_{0}^{c}}
\left\{\sup_{\|s-s_{{\rm obs}}\|\leq\delta'/3}f_{n}(s\mid\theta)\right\}+\bar{K}(\lambda_{\min}(\Lambda)\varepsilon_{n}^{-1}\delta'/3)\varepsilon_{n}^{-d},
\end{eqnarray*}
where $\lambda_{\min}(\Lambda)$ is positive.
In the above, as $n\rightarrow\infty$, the third
term is exponentially decreasing by Conditions \ref{kernel_prop}(iv). For the second term, by Condition \ref{sum_conv}, with probability $1$, 
\begin{align*}
\|s-s(\theta)\| & =\|\{s(\theta_0)-s(\theta)\}+\{s_{\rm obs}-s(\theta_0)\} +\varepsilon_n v\| \\
& \geq\delta'-\delta'/3-\delta'/3=\delta'/3.
\end{align*}
Recall that $W_{n}(s)=a_{n}A(\theta)^{-1/2}\{s-s(\theta)\}$. Then by Condition \ref{sum_approx_tail}, the second term is exponentially decreasing. For the first term, when $\theta\in B_{\delta}^{c}\backslash\mathcal{P}_{0}^{c}$ and $\|s-s_{{\rm obs}}\|\leq\delta'/3$, $\|W_{n}(s)\|\geq a_{n}\delta'r$
for some constant $r$. By Condition \ref{sum_approx} and \ref{sum_approx_tail}, $f_{W_{n}}(w\mid\theta)$ is bounded
by the sum of a normal density and  $\alpha_{n}^{-1}r_{{\rm max}}(w)$, which are both exponentially decreasing, so
$\sup_{\theta\in B_{\delta}^{c}\backslash\mathcal{P}_{0}^{c}}\sup_{\|s-s_{{\rm obs}}\|\leq\delta'/3}f_{n}(s\mid\theta)$
is also exponentially decreasing. Finally, the sum of all the above
is $O(e^{-a_{n,\varepsilon}^{\alpha_{\delta}}c_{\delta}})$ by noting
that $a_{n,\varepsilon}\leq\min(\varepsilon_{n}^{-1},a_{n})$. \end{proof}

The following additional lemma will be used repeatedly to bound error terms that appear in Lemmas \ref{posterior_normal} and \ref{lem:norm_remain}. 

\begin{lemma} \label{poly_mixnorm_bound} Assume Condition \ref{kernel_prop}.
For $t\in\mathbb{R}^{p}$ and $v\in\mathbb{R}^{d}$, let $\{A_{n}(t)\}$
be a series of $d\times p$ matrix functions, $\{C_{n}(t)\}$ be a
series of $d\times d$ matrix functions, $Q$ be a positive definite
matrix and $g_{1}(v)$ and $g_{2}(v)$ be probability densities in
$\mathbb{R}^{d}$. Let $c$ be a random vector, $\{k_{n}\}$ be
a series converging to $k_{\infty}\in(0,\infty]$ and $\{b_{n}'\}$
be a series converging to a non-negative constant. Let $b_{n}=\mathbbm{1}_{\{k_{\infty}=\infty\}}+b_{n}'\mathbbm{1}_{\{k_{\infty}<\infty\}}$.
If %the following conditions are satisfied,

(i) $g_{1}(v)$ and $g_{2}(v)$ are bounded in $\mathbb{R}^{d}$;

(ii) $g_{1}(v)$ and $g_{2}(v)$ depend on $v$ only through $\|v\|$
and are decreasing functions of $\|v\|$;

(iii) there exists an integer $l$ such that $\int\prod_{k=1}^{l+p}v_{i_{k}}g_{j}(v)\,dv<\infty$,
$j=1,2$, for any coordinates $(v_{i_{1}},\cdots,v_{i_{l}})$ of $v$;

(iv) there exists a positive constant $m$ such that for any $t\in\mathbb{R}^{p}$
and $n$, $\lambda_{{\rm min}}\{A_{n}(t)\}$ and $\lambda_{{\rm min}}\{C_{n}(t)\}$
are greater than $m$;

\noindent
then for any $P_{l}(t,v)\in\mathbb{P}_{l,(t,v)}$,
\begin{align*}
\int_{\mathbb{R}^{p}}\int_{\mathbb{R}^{d}}P_{l}(t,v)k_{n}^{d}g_{1}[k_{n}C_{n}(t)\{A_{n}(t)t-b_{n}v-k_{n}^{-1}c\}]g_{2}(Qv)\,dvdt & =O_p(1),\\
\int_{\mathbb{R}^{p}}\int_{\mathbb{R}^{d}}k_{n}^{d}g_{1}[k_{n}C_{n}(t)\{A_{n}(t)t-b_{n}v-k_{n}^{-1}c\}]g_{2}(Qv)\,dvdt & =\Theta_p(1).
\end{align*}
\end{lemma}

\begin{proof} For simplicity, here $\int$ denotes the integration over the whole Euclidean space. According to (ii), $g_{1}(v)$ can be written
as $\bar{g}_{1}(\|v\|)$. When $k_{\infty}<\infty$, assume $k_{n}=1$
without loss of generality. For any $P_{l}(t,v)\in\mathbb{P}_{l,(t,v)}$,
by Cauchy\textendash Schwarz inequality, there exists a $P_{l}(\|t\|,\|v\|)\in\mathbb{P}_{l,(\|t\|,\|v\|)}$
with coefficient functions taking positive values such that $|P_{l}(t,v)|$ is bounded
by $P_{l}(\|t\|,\|v\|)$ almost surely. Therefore for the first equality, it is
sufficient to consider the equality where $P_{l}(t,v)$ is replaced
by $P_{l}(\|t\|,\|v\|)$ and the coefficient functions of $P_{l}(\|t\|,\|v\|)$ are positive almost surely. For each $n$, divide $\mathbb{R}^{p}$ into $V=\{t:\|A_{n}(t)t\|/2\geq\|b_{n}'v+c\|\}$
and $V^{c}$. In $V$, $\|C_{n}(t)\{A_{n}(t)t-b_{n}'v-c\}\|\geq m^{2}\|t\|/2$;
in $V^{c}$, $\|t\|\leq2m^{-1}\|b_{n}'v+c\|$. With probability tending to $1$,
\begin{eqnarray*}
 \lefteqn{ \int P_{l}(\|t\|,\|v\|)g_{1}[C_n(t)\{A_n(t)t-b_{n}'v-c\}]g_{2}(Qv)\,dvdt \leq} \\
& & \int P_{l}(\|t\|,\|v\|)\bar{g}_{1}(m^{2}\|t\|/2)g_{2}(Qv)\,dvdt 
  +\sup_{v\in\mathbb{R}^{d}}g_{1}(v)\int\int_{V^{c}}dt\,P_{l}(2m^{-1}\|b_{n}'v+c\|,\|v\|)g_{2}(Qv)\,dv.
\end{eqnarray*}
In the above, $\int_{V^{c}}\,dt$ is the volume of $V^{c}$ in $\mathbb{R}^{p}$
and is proportional to $\|b_{n}'v+c\|^{p}$. By (iii), the right
hand side of the above inequality is $O_p(1)$. 

When $k_{\infty}=\infty$, let $v^{*}=k_{n}\{A(t)t-v-k_{n}^{-1}c\}$.
Then for any $P_{l}(t,v)\in\mathbb{P}_{l,(t,v)}$, with probability $1$,
\begin{align*}
 & \left|\int P_{l}(t,v)k_{n}^{d}g_{1}[k_{n}C_{n}(t)\{A(t)t-v-k_{n}^{-1}c\}]g_{2}(Qv)\,dvdt\right|\\
= & \left|\int P_{l}(t,v^{*})g_{2}[Q\{A(t)t-k_{n}^{-1}v^{*}-k_{n}^{-1}c\}]g_{1}(C_{n}(t)v^{*})\,dv^{*}dt\right|,\\
\leq & \int P_{l}(\|t\|,\|v^{*}\|)g_{2}[Q\{A(t)t-k_{n}^{-1}v^{*}-k_{n}^{-1}c\}]\overline{g}_{1}(m\|v^{*}\|)\,dv^{*}dt
\end{align*}
for some $P_{l}(t,v^{*})\in\mathbb{P}_{l,(t,v^{*})}$ and $P_{l}(\|t\|,\|v^{*}\|)\in\mathbb{P}_{l,(\|t\|,\|v^{*}\|)}$.
The right hand side of the above inequality is
similar to the integral when $k_{\infty}<\infty$ with $g_{1}(\cdot)$
and $g_{2}(\cdot)$ replaced by $g_{2}(\cdot)$ and $\overline{g}_{1}(\cdot)$
respectively. Therefore it is $O_p(1)$ by the same reasoning.

For $P_{l}(t,v)=1$, by considering only the integral in a compact
region, it is easy to see the target integral is larger than $0$.
Therefore the lemma holds. \end{proof}

\begin{proof}[of Lemma \ref{t_mixnorm_val}] Let $P=A^{T}A$. By matrix algebra, 
\[
N\Big(At;B_{n}v+\frac{1}{k_{n}}c,\frac{1}{k_{n}^{2}}I_{d}\Big)K(v)=N\Big\{ t;P^{-1}A^{T}\left(B_{n}v+\frac{1}{k_{n}}c\right),\frac{1}{k_{n}^{2}}P^{-1}\Big\} r(v;A,B_{n,}k_{n},c),
\]
where 
\begin{align*}
r(v;A,B_{n,}k_{n},c) & =\frac{k_{n}^{d-p}}{(2\pi)^{(d-p)/2}}\exp\Big\{-\frac{k_{n}^{2}}{2}\left(B_{n}v+\frac{c}{k_{n}}\right)^{T}(I-AP^{-1}A^{T})\left(B_{n}v+\frac{c}{k_{n}}\right)\Big\} K(v).
\end{align*}
Then the target integral can be expanded as 
\begin{align*}
\int t\frac{N(At;B_{n}v+\frac{1}{k_{n}}c,\frac{1}{k_{n}^{2}}I_{d})K(v)}{\int N(At;B_{n}v+\frac{1}{k_{n}}c,\frac{1}{k_{n}^{2}}I_{d})K(v)\,dtdv}\,dtdv & 
=\int P^{-1}A^{T}\left(\frac{1}{k_{n}}c+B_{n}v\right)\frac{r(v;A,B_{n,}k_{n},c)}{\int r(v;A,B_{n,}k_{n},c)\,dv}\,dv\\
 & =\frac{1}{k_{n}}\left\{ (A^{T}A)^{-1}A^{T}c+R(A,B_{n},k_{n},c)\right\} ,
\end{align*}
where
\[
R(A,B_{n},k_{n},c)=(A^{T}A)^{-1}A^{T}B_{n}\int k_{n}v\frac{r(v;A,B_{n,}k_{n},c)}{\int r(v;A,B_{n,}k_{n},c)\,dv}\,dv.
\]

The remainder term $R(A,B_{n},k_{n},c)$ depends on the mean of the probability density proportional
to $r(v;A,B_{n,}k_{n},c)$ in the directions of $(A^{T}A)^{-1}A^{T}B$.
If $B_{n}$ does not degenerate to $0$ as $n\rightarrow\infty$,
then in the directions orthogonal to those of $(I-A(A^{T}A)^{-1}A^{T})^{1/2}B$,
$r(v;A,B_{n,}k_{n},c)$ is symmetric around $0$; in the directions
of $(I-A(A^{T}A)^{-1}A^{T})^{1/2}B$, $r(v;A,B_{n,}k_{n},c)$ is a
product of a normal density whose mean is $O(1/k_{n})$ and
a rescaled $K(v)$, which is symmetric around $0$, so its mean value is $O(1/k_{n})$. Therefore when the spaces expanded by $(A^{T}A)^{-1}A^{T}B$
and $\{I-A(A^{T}A)^{-1}A^{T}\}B$ are orthogonal, $R(A,B_{n},k_{n},c)=0$;
when it is not the case, $R(A,B_{n},k_{n},c)=O(1)$. 

If $B_{n}=o(1)$ as $n\rightarrow\infty$, which implies $k_{n}\rightarrow c\in(0,\infty)$, it is easy to see that $\int k_{n}vr(v;A,B_{n,}k_{n},c)\,dv/\int r(v;A,B_{n,}k_{n},c)\,dv$
is upper bounded as $n\rightarrow\infty$ and hence $R(A,B_{n},k_{n},c)$
is $o(1)$. 
\end{proof}

In the following lemmas, to deal with the case where $K(x)=\bar{K}(||x||_{\Lambda})$ with $\Lambda$ not the identity, we use the property that such a $K(x)$ can be bounded above by a function that depends 
only on $||x||$. We refer to this bound as $K(\cdot)$ rescaled to have identity covariance matrix.

\begin{proof}[of Lemma \ref{posterior_normal}] First consider $\tpi_{B_{\delta}}(1)$. With the transformation
$t=t(\theta)$, 
\begin{align}
\tpi_{B_{\delta}}(1) & =a_{n,\varepsilon}^{-p}\int_{t(B_{\delta})}\int_{\mathbb{R}^{d}}\pi(\theta_{0}+a_{n,\varepsilon}^{-1}t)\ftil_{n}(s_{{\rm obs}}+\varepsilon_{n}v\mid\theta_{0}+a_{n,\varepsilon}^{-1}t)K(v)\,dvdt.\label{norm_expand1}
\end{align}
We can obtain an expansion of $\tpi_{B_{\delta}}(1)$ by expanding
$\ftil_{n}(s_{{\rm obs}}+\varepsilon_{n}v\mid\theta_{0}+a_{n,\varepsilon}^{-1}t)K(v)$
as follows. The expansion needs to be discussed separately for two
cases, depending on whether the limit of $a_{n}\varepsilon_{n}$ is
finite or infinite.

When $a_{n}\varepsilon_{n}\rightarrow c_{\varepsilon}<\infty$, $a_{n,\varepsilon}=a_{n}$.
We apply a Taylor expansion to $s(\theta_{0}+a_{n}^{-1}t)$ and $A(\theta_{0}+a_{n}^{-1}t)^{-1/2}$ and have 
\begin{align}
 & \ftil_{n}(s_{{\rm obs}}+\varepsilon_{n}v\mid\theta_{0}+a_{n}^{-1}t)= \frac{a_{n}^{d}}{|A(\theta_{0}+a_{n}^{-1}t)|^{1/2}}\nonumber \\
\times & N\left(\Big\{ A(\theta_{0})^{-1/2}+a_{n}^{-1}r_{A}(t,\epsilon_{2})\Big\}\Big[A(\theta_{0})^{1/2}T_{\rm obs}+a_{n}\varepsilon_{n}v-\{Ds(\theta_{0})+a_{n}^{-1}r_{s}(t,\epsilon_{1})\}t\Big];0,I_{d}\right),\label{eq:f_til_expand1}
\end{align}
where $r_{s}(t,\epsilon_{1})$ is the $d\times p$ matrix whose
$i$th row is $t^{T}Hs_{i}\{\theta_{0}+\epsilon_{1}(t)\}$, $r_{A}(t,\epsilon_{2})$
is the $d\times d$ matrix $\sum_{k=1}^{p}\frac{d}{d\theta_{k}}A\{\theta_{0}+\epsilon_{2}(t)\}^{-1/2}t_{k}$,
and $\epsilon_{1}(t)$ and $\epsilon_{2}(t)$ are from the remainder terms
of the Taylor expansions and satisfy $\|\epsilon_{1}(t)\|\leq\delta$ and
$\|\epsilon_{2}(t)\|\leq\delta$. For a $d\times d$ matrix $\tau_2$, let $g_n(t,v;\tau_1,\tau_2)$ be the function $g_n(t,v;\tau_1)$, defined in Section \ref{sec:proposal} of the main text, with $A(\theta_0)$ replaced by $\{A(\theta_0)^{-1/2}+\tau_2\}^{-2}$. Applying a Taylor expansion to
the normal density in \eqref{eq:f_til_expand1}, we have 
\begin{align}
 & \ftil_{n}(s_{{\rm obs}}+\varepsilon_{n}v\mid\theta_{0}+a_{n}^{-1}t)K(v)\nonumber \\
= & \frac{a_{n}^{d}|A(\theta_{0})|^{1/2}}{|A(\theta_{0}+a_{n}^{-1}t)|^{1/2}}
\Big[g_{n}(t,v)+a_{n}^{-1}P_{3}(t,v)g_{n}\{t,v;e_{n1}r_{s}(t,\epsilon_{1}),e_{n1}r_{A}(t,\epsilon_{2})\}\Big],\label{eq:fK_expand1}
\end{align}
where $P_{3}(t,v)$ is the function 
\begin{eqnarray*}
\lefteqn{\frac{1}{2|A(\theta_{0})^{-1/2}+r_{2}(a_{n}^{-1}t)|}} \\
&\times& \left.\frac{d}{dx}\left\Vert \left\{ A(\theta_{0})^{-1/2}+xr_{A}(t,\epsilon_{2})\right\} \left[A(\theta_{0})^{1/2}T_{\rm obs}+a_{n}\varepsilon_{n}v-\{Ds(\theta_{0})+xr_{s}(t,\epsilon_{1})\}t\right]\right\Vert ^{2}\right|_{x=e_{n1}},
\end{eqnarray*}
and $e_{n1}$ is from the remainder term of Taylor expansion and satisfies
$|e_{n1}|\leq a_{n}^{-1}$ . Since $\|e_{n1}t\|\leq\delta$ and $r_{s}(t,\epsilon_{1})$
and $r_{A}(t,\epsilon_{2})$ belong $\mathbb{P}_{1,t}$, this $P_{3}(t,v)$
belongs to $\mathbb{P}_{3,(t,v)}$. Furthermore, since $r_{s}(t,\epsilon_{1})$
and $r_{A}(t,\epsilon_{2})$ have no constant term, for any small
$\sigma$, $e_{n1}r_{s}(t,\epsilon_{1})$ and $e_{n1}r_{A}(t,\epsilon_{2})$
can be bounded by $\sigma I_{d}$ and $\sigma I_{p}$ uniformly in
$n$ and $t$, if $\delta$ is small enough. 

When $a_{n}\varepsilon_{n}\rightarrow\infty$, $a_{n,\varepsilon}=\varepsilon_{n}^{-1}$.
Let $v^{*}(v)=A(\theta_{0})^{1/2}T_{{\rm obs}}+a_{n}\varepsilon_{n}v-a_{n}\varepsilon_{n}Ds(\theta_{0})t$.
Under the transformation $v^{*}=v^{*}(v)$, the expansion of $\ftil_{n}(s_{{\rm obs}}+\varepsilon_{n}v\mid\theta_{0}+\varepsilon_{n}t)$
%similar to \eqref{eq:f_til_expand1}, 
obtained by applying a Taylor expansion
to $s(\theta_{0}+\varepsilon_{n}t)$ and $A(\theta_{0}+\varepsilon_{n}t)^{-1/2}$
is 
\begin{align*}
 & \ftil_{n}(s_{{\rm obs}}+\varepsilon_{n}v\mid\theta_{0}+\varepsilon_{n}t)\\
= & \frac{a_{n}^{d}}{|A(\theta_{0}+a_{n}^{-1}t)|^{1/2}}N\left[\Big\{ A(\theta_{0})^{-1/2}+a_{n}\varepsilon_{n}^{2}\frac{r_{A}(t,\epsilon_{4})}{a_{n}\varepsilon_{n}}\Big\}
\Big\{ v^{*}-a_{n}\varepsilon_{n}^{2}r_{s}(t,\epsilon_{3})t\Big\};0,I_{d}\right],
\end{align*}
where $\epsilon_{3}(t)$ and $\epsilon_{4}(t)$ are from the remainder terms
of the Taylor expansion and satisfy $\|\epsilon_{3}(t)\|\leq\delta$ and
$\|\epsilon_{4}(t)\|\leq\delta$. Let $g_{n}^{*}(t,v^{*};\tau_{1},\tau_{2})$
be the function 
\begin{eqnarray*}
\lefteqn{g_{n}^{*}(t,v^{*};\tau_{1},\tau_{2})}\\
& = &N\left[v^{*};a_{n}\varepsilon_{n}\tau_{1}t,\{A(\theta_{0})^{-1/2}+\tau_{2}\}^{-2}\right]
K\left\{Ds(\theta_{0})t+\frac{1}{a_{n}\varepsilon_{n}}v^{*}-\frac{1}{a_{n}\varepsilon_{n}}A(\theta_{0})^{1/2}T_{\rm obs}\right\},
\end{eqnarray*}
so that $(a_{n}\varepsilon_{n})^{d}g_{n}^{*}(t,v^{*};\tau_{1},\tau_{2})$
is $g_{n}(t,v;\tau_{1},\tau_{2})$ with transformed variable $v^{*}=v^{*}(v)$,
and $g_{n}^{*}(t,v^{*})=g_{n}^{*}(t,v^{*};0,0)$. Denote a $k_{1}\times k_{2}$
matrix with element being $P_{l}(t)$ by $P_{l}^{(k_{1}\times k_{2})}(t)$.
Then by applying a Taylor expansion to the normal density in the expansion
above,  
\begin{align}
 & \ftil_{n}(s_{{\rm obs}}+\varepsilon_{n}v\mid\theta_{0}+\varepsilon_{n}t)K(v)\nonumber \\
= & \frac{\varepsilon_{n}^{-d}|A(\theta_{0})|^{1/2}}{|A(\theta_{0}+\varepsilon_{n}t)|^{1/2}}\Big[g_{n}^{*}(t,v^{*})+a_{n}\varepsilon_{n}^{2}\left\{ P_{2}^{(d\times1)}(t)v^{*}+\frac{1}{a_{n}\varepsilon_{n}}v^{*T}P_{1}^{(d\times d)}(t)v^{*}\right\} g_{n}^{*}(t,v^{*})\nonumber \\
 & +(a_{n}\varepsilon_{n}^{2})^{2}P_{4}(t,v^{*})g_{n}^{*}\{t,v^{*};e_{n2}r_{s}(t,\epsilon_{3}),e_{n2}r_{A}(t,\epsilon_{4})\}\Big](a_{n}\varepsilon_{n})^{d},\label{eq:fK_expand2}
\end{align}
where $P_{2}^{(d\times1)}(t)$ is the function $t^{T}r_{s}(t,\epsilon_{3})^{T}A(\theta_{0})^{-1/2}/2$,
$P_{1}^{(d\times d)}(t)$ is the function $-A(\theta_{0})^{-1/2}r_{A}(t,\epsilon_{4})$,
$e_{n2}=e_{n2}'/(a_{n}\varepsilon_{n})$, $e_{n2}'$ is from the remainder term
of the Taylor expansion and satisfies $|e_{n2}'|\leq a_{n}\varepsilon_{n}^{2}$,
and $P_{4}(t,v^{*})$ is a linear combination of $\{d\rho(w)/dw\}^{2}$
and $d^{2}\rho(w)/dw^{2}$ at $w=e_{n2}'$ with $\rho(w)$ being the
function 
\[
\Big\|\Big\{ A(\theta_{0})^{-1/2}+w\frac{r_{A}(t,\epsilon_{4})}{a_{n}\varepsilon_{n}}\Big\}\Big\{ v^{*}-wr_{s}(t,\epsilon_{3})t\Big\}\Big\|^{2}.
\]
Obviously elements of $P_{2}^{(d\times1)}(t)$ and $P_{1}^{(d\times d)}(t)$
belong to $\mathbb{P}_{2,t}$ and $\mathbb{P}_{1,t}$ respectively.
Since $\|e_{n2}t\|\leq\delta$, the function $P_{4}(t,v^{*})$ belongs to $\mathbb{P}_{4,(t,v^{*})}$
and, similar to before, $e_{n2}r_{s}(t,\epsilon_{3})$ and $e_{n2}r_{A}(t,\epsilon_{4})$
can be bounded by $\sigma I_{d}$ and $\sigma I_{p}$ uniformly in
$n$ and $t$ for any small $\sigma$, if $\delta$ is small enough. 

For $\pi(\theta_{0}+a_{n,\varepsilon}^{-1}t)$ in the integral of
$\tpi_{B_{\delta}}(1)$ in \eqref{norm_expand1}, a Taylor expansion
gives that 
\begin{align}
\frac{\pi(\theta_{0}+a_{n,\varepsilon}^{-1}t)}{|A(\theta_{0}+a_{n,\varepsilon}^{-1}t)|^{1/2}} & =
\frac{\pi(\theta_{0})}{|A(\theta_{0})|^{1/2}}+
a_{n,\varepsilon}^{-1}D_{\theta}\frac{\pi\{\theta_{0}+\epsilon_{5}(t)\}}{|A\{\theta_{0}+\epsilon_{5}(t)\}|^{1/2}}t,\quad |\epsilon_{5}(t)|\leq\delta.\label{eq:fK_expand3}
\end{align}

As mentioned before, $\delta$ can be selected such that $Ds(\theta_{0})+e_{n1}r_{s}(t,\epsilon_{1})$
and $Ds(\theta_{0})+e_{n2}r_{s}(t,\epsilon_{3})$ are lower bounded
by $m_{1}I_{p}$ and $A(\theta_{0})^{-1/2}+e_{n1}r_{A}(t,\epsilon_{2})$
and $A(\theta_{0})^{-1/2}+e_{n2}r_{A}(t,\epsilon_{4})$ are lowered
bounded by $m_{2}I_{d}$ for some positive constant $m_{1}$ and $m_{2}$.
We choose $\delta$ satisfying these and, since $\|a_{n,\varepsilon}^{-1}t\|\leq\delta$,
this means $\pi(\theta_{0}+a_{n,\varepsilon}^{-1}t)/|A(\theta_{0}+a_{n,\varepsilon}^{-1}t)|^{1/2}$
is bounded uniformly in $t$ and $n$. 

By plugging \eqref{eq:fK_expand1}\textendash \eqref{eq:fK_expand3}
into \eqref{norm_expand1}, it can be seen that the leading term of
$\tpi_{B_{\delta}}(1)$ is $a_{n,\varepsilon}^{d-p}\pi(\theta_{0})\int_{t(B_{\delta})\times\mathbb{R}^{d}}g_{n}(t,v)\,dtdv$.
The remainder terms are given in the following, 
\begin{align}
 & a_{n,\varepsilon}^{p-d}\tpi_{B_{\delta}}(1)-\pi(\theta_{0})\int_{t(B_{\delta})\times\mathbb{R}^{d}}g_{n}(t,v)\,dtdv\nonumber \\
= & a_{n,\varepsilon}^{-1}\int_{t(B_{\delta})\times\mathbb{R}^{d}}|A(\theta_{0})|^{1/2}D\frac{\pi(\theta_{0}+\epsilon_{5})}{|A(\theta_{0}+\epsilon_{5})|^{1/2}}tg_{n}(t,v)\,dvdt\nonumber \\
 & +a_{n}^{-1}\int_{t(B_{\delta})\times\mathbb{R}^{d}}P_{3}(t,v)g_{n}\{t,v;e_{n1}r_{s}(t,\epsilon_{1}),e_{n1}r_{A}(t,\epsilon_{2})\}\,dvdt\ \mathbbm{1}_{\{\lim a_{n}\varepsilon_{n}<\infty\}}\nonumber \\
 & +a_{n}\varepsilon_{n}^{2}\int_{t(B_{\delta})}P_{2}^{(d\times1)}(t)\int_{\mathbb{R}^{d}}v^{*}g_{n}^{*}(t,v^{*})\,dv^{*}dt\ \mathbbm{1}_{\{\lim a_{n}\varepsilon_{n}=\infty\}}\nonumber \\
 & +\varepsilon_{n}\int_{t(B_{\delta})\times\mathbb{R}^{d}}v^{*T}P_{1}^{(d\times d)}(t)v^{*}g_{n}^{*}(t,v^{*})\,dv^{*}dt\ \mathbbm{1}_{\{\lim a_{n}\varepsilon_{n}=\infty\}}\nonumber \\
 & +a_{n}^{2}\varepsilon_{n}^{4}\int_{t(B_{\delta})\times\mathbb{R}^{d}}P_{4}(t,v^{*})g_{n}^{*}\{t,v^{*};e_{n2}r_{s}(t,\epsilon_{3}),e_{n2}r_{A}(t,\epsilon_{4})\}\,dv^{*}dt\ \mathbbm{1}_{\{\lim a_{n}\varepsilon_{n}=\infty\}},\label{norm_expand3}
\end{align}
where $P_{3}(t,v)$, $P_{2}^{(d\times1)}(t)$, $P_{1}^{(d\times d)}(t)$
and $P_{4}(t,v^{*})$ are products of $\pi(\theta_{0}+a_{n,\varepsilon}^{-1}t)/|A(\theta_{0}+a_{n,\varepsilon}^{-1}t)|^{1/2}$
and corresponding terms in expansions \eqref{eq:fK_expand1} and \eqref{eq:fK_expand2}.
In the above, there are five remainder terms. For the integrals in
the first two terms, it is easy to write them in the form of the first
integral in Lemma \ref{poly_mixnorm_bound} and conditions therein
are satisfied, where $g_{1}(\cdot)$ is the standard normal density
and $g_{2}(\cdot)$ is $K(v)$ rescaled to have identity covariance.
Then the first two terms are $O_{p}(a_{n,\varepsilon}^{-1})$ and
$O_{p}(a_{n}^{-1})$. The integral in the fourth term  can
also be written in this form where $g_{1}(\cdot)$ is the rescaled
$K(v)$ and $g_{2}(\cdot)$ is the standard normal density. The
integral in the fifth term needs to use the transformation $v^{**}=v^{*}-a_{n}\varepsilon_{n}e_{n2}r_{s}(t,\epsilon_{3})t$, after which it can
 be written in a similar form, as $P_{5}\{t,v^{**}+a_{n}\varepsilon_{n}e_{n2}r_{s}(t,\epsilon_{3})t\}\in\mathbb{P}_{5,(t,v^{**})}$
by the expression of $P_{4}(t,v^{*})$ in \eqref{eq:fK_expand2}.
Thus the fourth and fifth term are $O_{p}(\varepsilon_{n})$ and $O_{p}(a_{n}^{2}\varepsilon_{n}^{4})$.

The third term is somewhat different as the center of $g_{n}^{*}(t,v^{*})$
in the direction of $v^{*}$ degenerates to zero as $n\rightarrow\infty$. Let $\psi_{k}$ be the
$d$-dimension unit vector with $1$ at the $k$th coordinate. Then  
\begin{align*}
\int_{-\infty}^{\infty}v_{k}^{*}g_{n}^{*}(t,v^{*})\,dv_{k}^{*} & =\int_{0}^{\infty}v_{k}^{*}\{g_{n}^{*}(t,v^{*})-g_{n}^{*}(t,v^{*}-2v_{k}^{*}\psi_{k})\}\,dv_{k}^{*}\\
 & =\int_{0}^{\infty}v_{k}^{*}N\{v^{*};0,A(\theta_{0})\}[K\{v(v^{*})\}-K\{v(v^{*}-2v_{k}^{*}\psi_{k})\}]\,dv_{k}^{*},
\end{align*}
which by a Taylor expansion is bounded by $(a_{n}\varepsilon_{n})^{-1}c$
for some constant $c$. Hence the third term is  $O_{p}(\varepsilon_{n})$.
Combining the orders of all remainder terms, the expansion
of $\tpi_{B_{\delta}}(1)$ in the lemma holds. 

For any $P_{2}(t)\in\mathbb{P}_{2,t}$, $\widetilde{\pi}_{B_{\delta}}\{P_{2}(t)\}$
can be expanded similarly to $\tpi_{B_{\delta}}(1)$ in \eqref{norm_expand3},
simply by multplying $P_{2}(t)$ into every integral in \eqref{norm_expand3}.
This gives that 
\[
\tpi_{B_{\delta}}\{P_{2}(t)\}=a_{n,\varepsilon}^{d-p}\Big\{\pi(\theta_{0})\int_{t(B_{\delta})\times\mathbb{R}^{d}}P_{2}(t)g_{n}(t,v)\,dtdv+O_{p}(a_{n,\varepsilon}^{-1})+O_{p}(a_{n}^{2}\varepsilon_{n}^{4})\Big\}.
\]
Then since $\int_{t(B_{\delta})\times\mathbb{R}^{d}}g_{n}(t,v)\,dtdv=\Theta_{p}(1)$  by the second result of Lemma \ref{poly_mixnorm_bound},
$\tpi_{B_{\delta}}\{P_{2}(t)\}/\tpi_{B_{\delta}}(1)=O_{p}(1)$ and \eqref{eq:post_lim2}
holds by taking $P_{2}(t)=t$. \end{proof}

\begin{proof}[of Lemma \ref{lem:norm_remain}]
%The error of $\widetilde{\pi}_{B_{\delta}}(P_{l}(t))$
%to approximate $\pi_{B_{\delta}}(P_{l}(t))$ for $P_{l}(t)\in\mathbb{P}_{l,t}$
%and some $l\geq2$. 
Let $r_{n}(s\mid\theta)$ be the scaled remainder
$\alpha_{n}\{f_{n}(s\mid\theta)-\ftil_{n}(s\mid\theta)\}$. The
error of using $\widetilde{\pi}_{B_{\delta}}\{P_{l}(t)\}$ to approximate $\pi_{B_{\delta}}\{P_{l}(t)\}$ is
\[
\pi_{B_{\delta}}\{P_{l}(t)\}-\tpi_{B_{\delta}}\{P_{l}(t)\}=\alpha_{n}^{-1}\int_{B_{\delta}}\int P_{l}\{t(\theta)\}\pi(\theta)r_{n}(s_{{\rm obs}}+\varepsilon_{n}v\mid\theta)K(v)\,dvd\theta.
\]
If this approximation error satisfies 
\begin{align}
\frac{\pi_{B_{\delta}}\{P_{l}(t)\}-\tpi_{B_{\delta}}\{P_{l}(t)\}}{\tpi_{B_{\delta}}(1)} & =O_{p}(\alpha_{n}^{-1}),\label{eq:norm_remain}
\end{align}
then, since $a_{n,\varepsilon}^{p-d}\tpi_{B_{\delta}}(1)=\Theta_{p}(1)$
 by Lemma \ref{posterior_normal}, 
\begin{align}
\pi_{B_{\delta}}(1)=\tpi_{B_{\delta}}(1)\{1+O_{p}(\alpha_{n}^{-1})\}, \quad
\frac{\pi_{B_{\delta}}\{P_{l}(t)\}}{\pi_{B_{\delta}}(1)} & =\frac{\tpi_{B_{\delta}}\{P_{l}(t)\}}{\tpi_{B_{\delta}}(1)}+O_{p}(\alpha_{n}^{-1}).\label{norm_approx}
\end{align}
By plugging \eqref{eq:norm_remain} into \eqref{h_expand}, 
\begin{equation}
\frac{\pi_{B_{\delta}}(h)}{\pi_{B_{\delta}}(1)}=h(\theta_{0})+a_{n,\varepsilon}^{-1}Dh(\theta_{0})^{T}\left\{\frac{\tpi_{B_{\delta}}(t)}{\tpi_{B_{\delta}}(1)}+O_{p}(\alpha_{n}^{-1})\right\}+
\frac{1}{2}a_{n,\varepsilon}^{-2}\left[\frac{\tpi_{B_{\delta}}\{t^{T}Hh(\theta_{t})t\}}{\tpi_{B_{\delta}}(1)}+O_{p}(\alpha_{n}^{-1})\right].\label{norm_expand7}
\end{equation}

Verification of \eqref{eq:norm_remain} is given by the following argument. With the transformation $t=t(\theta)$ we have
\begin{align*}
\pi_{B_{\delta}}\{P_{l}(t)\}-\tpi_{B_{\delta}}\{P_{l}(t)\} & =\alpha_{n}^{-1}a_{n,\varepsilon}^{-p}\int_{t(B_{\delta})}\int P_{l}(t)\pi(\theta_{0}+a_{n,\varepsilon}^{-1}t)r_{n}(s_{{\rm obs}}+\varepsilon_{n}v\mid\theta_{0}+a_{n,\varepsilon}^{-1}t)K(v)\,dvdt.
\end{align*}
Let $r_{W_{n}}(w\mid\theta)=\alpha_{n}\{f_{W_{n}}(w\mid\theta)-\widetilde{f}_{W_{n}}(w\mid\theta)\}$,
and we have \[r_{n}(s\mid\theta)=a_{n}^{d}\vert A(\theta)\vert^{-1/2}r_{W_{n}}[a_{n}A(\theta)^{-1/2}\{s-s(\theta)\}\mid\theta].\]
For the value of $\delta$, we choose the smaller value of the
one from Lemma \ref{posterior_normal} and the one such that
$Ds(\theta)$ is lower bounded and $A(\theta)^{-1/2}$ is upper bounded
by $MI_{d}$ in $B_{\delta}$ for some $M>0$. Since $r_{W_{n}}(w\mid\theta)$
is upper bounded by $r_{\rm max}(w)$ according to Condition \ref{sum_approx},
by applying a Taylor expansion to $s(\theta_{0}+a_{n,\varepsilon}^{-1}t)$ we have 
\begin{align*}
 & |\pi_{B_{\delta}}\{P_{l}(t)\}-\tpi_{B_{\delta}}\{P_{l}(t)\}|\leq\alpha_{n}^{-1}a_{n,\varepsilon}^{d-p}\sup_{\theta\in B_{\delta}}\vert\pi(\theta)A(\theta)^{-1/2}\vert \int_{t(B_{\delta})}\int 
 |P_{l}(t)|(a_{n}a_{n,\varepsilon}^{-1})^{d} \\
 & r_{{\rm max}}\Big[
 a_{n}a_{n,\varepsilon}^{-1}M\Big\{ Ds(\theta_{0}+\epsilon_t)t-a_{n,\varepsilon}\varepsilon_{n}v-\frac{1}{a_{n}a_{n,\varepsilon}^{-1}}A(\theta_{0})^{1/2}T_{\rm obs}\Big\}\Big]K(v)\,dvdt,
\end{align*}
where $\epsilon_t$ is from the remainder term of the Taylor expansion and satisfies
$\vert\epsilon_t\vert\leq\delta$. Since $\tpi_{B_{\delta}}(1)=\Theta_{p}(a_{n,\varepsilon}^{d-p})$
by Lemma \ref{posterior_normal}, it is sufficient to show that the
above integral is $O_{p}(1)$. This is immediate by noting that when
either $\lim a_{n}\varepsilon_{n}\rightarrow\infty$ or $\lim a_{n}\varepsilon_{n}\rightarrow c_{\varepsilon}<\infty$,
the above integral can be written in the form of the first integral
in Lemma \ref{poly_mixnorm_bound} and conditions therein are satisfied,
where $g_{1}(\cdot)$ and $g_{2}(\cdot)$ are $r_{\rm max}(\cdot)$ and
$K(\cdot)$ rescaled to have identity covariance matrix. 
\end{proof}

\section{Proof of Results from Section 4}

\label{App:B}

\subsection{Proof of Proposition \ref{prop:Monte_Carlo_CLT}}

The proof of Proposition \ref{prop:Monte_Carlo_CLT} follows the standard
asymptotic argument of importance sampling. In the following we use
the convention that for a vector $x$, the matrix $xx^{T}$
is denoted by $x^{2}$. 
\begin{proof}[of Proposition \ref{prop:Monte_Carlo_CLT}]
Algorithm 1 generates independent, indentically distributed triples,
$(\phi_{i},\theta_{i},s_{n}^{(i)})$, where $(\theta_{i},s_{n}^{(i)})$
is generated from $g_{n}(\theta)f(s_{n}\mid\theta)$, and, conditional
on $s_{n}=s_{n}^{(i)}$, $\phi_{i}$ is generated from a Bernoulli
distribution with probability $K_{\varepsilon_{n}}(s_{n}-s_{{\rm obs}})$.

Now $\hat{h}$ can be expressed as a ratio of sample means of functions
of these independent, indentically distributed random variables. Thus
we can use the standard delta method \cite[]{lehmann2004elements}
for ratio statistics to show that the central limit theorem holds.
Further we obtain that the limiting distribution has mean 
\begin{align*}
\frac{E\{h(\theta_{1})w_{1}\phi_{1}\}}{E(w_{1}\phi_{1})}=
\frac{E\{h(\theta_{1})w_{1}K_{\varepsilon_{n}}(s_{n}^{(1)}-s_{{\rm obs}})\}}{E\{w_{1}K_{\varepsilon_{n}}(s_{n}^{(1)}-s_{{\rm obs}})\}}=
\frac{\int h(\theta)\pi(\theta)f_{n}(s_{n}\mid\theta)K_{\varepsilon_{n}}(s_{n}-s_{{\rm obs}})\,ds_{n}\,d\theta}{\int\pi(\theta)f_{n}(s_{n}\mid\theta)K_{\varepsilon_{n}}(s_{n}-s_{{\rm obs}})\,ds_{n}\,d\theta},%=h_{\rm ABC},
\end{align*}
which is equal to $h_{\rm ABC}$.
Its variance is
\begin{align*}
 & \frac{1}{E^{2}(w_{1}\phi_{1})}\mbox{var}\{h(\theta_{1})w_{1}\phi_{1}\}+
 \frac{E^{2}\{h(\theta_{1})w_{1}\phi_{1}\}}{E^{4}(w_{1}\phi_{1})}\mbox{var}(w_{1}\phi_{1})-2\frac{E\{h(\theta_{1})w_{1}\phi_{1}\}}{E^{3}(w_{1}\phi_{1})}
 \mbox{cov}\{h(\theta_{1})w_{1}\phi_{1},w_{1}\phi_{1}\}^{T}\\
= & p_{{\rm acc},\pi}^{-2}\left[
E\{h(\theta_{1})^{2}w_{1}^{2}\phi_{1}\}-h_{\rm ABC}^{2}p_{{\rm acc},\pi}^{2}+h_{\rm ABC}^{2}\left\{ E(w_{1}^{2}\phi_{1})-p_{{\rm acc},\pi}^{2}\right\} \right.\\
&\left. -2h_{\rm ABC}\left\{ E\{h(\theta_{1})w_{1}^{2}\phi_{1}\}-h_{\rm ABC}p_{{\rm acc},\pi}^{2}\right\} ^{T}\right]\\
= & p_{{\rm acc},\pi}^{-2}E[\{h(\theta_{1})^{2}-2h_{\rm ABC}h(\theta_{1})^{T}+h_{\rm ABC}^{2}\}w_{1}^{2}K_{\varepsilon_{n}}(s_{n}^{(1)}-s_{{\rm obs}})]\\
= & p_{{\rm acc},\pi}^{-1}E_{\pi_{\rm ABC}}\left\{(h(\theta)-h_{\rm ABC})^{2}\frac{\pi(\theta)}{q_{n}(\theta)}\right\}.
 \end{align*}
In the above expression we used  $p_{{\rm acc},\pi}=E(w_{1}\phi_{1})$.
It is easy to verify that 
\begin{equation}
\Sigma_{{\rm ABC},n}=p_{{\rm acc},\pi}^{-1}E_{\pi_{\rm ABC}}\left\{(h(\theta)-h_{\rm ABC})^{2}\frac{\pi(\theta)}{q_{n}(\theta)}\right\},\label{eq:expression_Sigma_ABC1}
\end{equation}
as required. \end{proof}

\subsection{Proof of Theorem 2}

For simplicity, a consider one-dimensional function $h(\theta)$. For multi-dimensional functions, the extension is trivial by considering each element of $\Sigma_{{\rm IS},n}$ seperately. Denote $\{h(\theta)-h_{\rm ABC}\}^{2}$
by $G_{n}(\theta)$. In Theorem \ref{thm:2}(i), $\Sigma_{{\rm IS},n}$
is just the ABC posterior variance of $h(\theta)$, and the derivation
of its order is similar to that of $h_{\rm ABC}$ in Section \ref{App:A} of this supplementary material.
The result is stated in the following lemma.

\begin{lemma} \label{ABCvar} Assume the conditions of Theorem \ref{thm:1}.
Then $\mbox{var}_{\pi_{\rm ABC}}\{h(\theta)\}=O_{p}(a_{n,\varepsilon}^{-2})$.
\end{lemma}

\begin{proof} Using the notation of Section \ref{App:A}, $\mbox{var}_{\pi_{\rm ABC}}[h(\theta)]=\pi(G_{n})/\pi(1)$.
It follows immediately from Lemma \ref{marglik_ignored} that 
\[
\mbox{var}_{\pi_{\rm ABC}}\{h(\theta)\}=\frac{\pi_{B_{\delta}}(G_{n})}{\pi_{B_{\delta}}(1)}\{1+o_{p}(1)\}.
\]
Applying a first order Taylor expansion of $h(\theta)$ around
$\theta=\theta_{0}$ gives 
\begin{equation}
\frac{\pi_{B_{\delta}}(G_{n})}{\pi_{B_{\delta}}(1)}=G_{n}(\theta_{0})+2a_{n,\varepsilon}^{-1}\{h(\theta_{0})-h_{\rm ABC}\}
\frac{\pi_{B_{\delta}}\{Dh(\theta_{t})^{T}t\}}{\pi_{B_{\delta}}(1)}+a_{n,\varepsilon}^{-2}\frac{\pi_{B_{\delta}}\{t^{T}Dh(\theta_{t})Dh(\theta_{t})^{T}t\}}{\pi_{B_{\delta}}(1)},\label{eq:expansion_variance}
\end{equation}
where $\theta_{t}$ is from the remainder term and belongs to $B_{\delta}$.
In the above decomposition, $G_{n}(\theta_{0})$ and $a_{n,\varepsilon}^{-1}\{h(\theta_{0})-h_{\rm ABC}\}$
are $O_{p}(a_{n,\varepsilon}^{-2})$ by Theorem \ref{thm:1}. Since
$Dh(\theta_{t})^{T}t$ and $t^{T}Dh(\theta_{t})Dh(\theta_{t})^{T}t$
belong to $\mathbb{P}_{2,t}$, the two ratios in the above are $O_{p}(1)$
by Lemma \ref{posterior_normal} and Lemma \ref{lem:norm_remain}.
%Thus the lemma holds.
\end{proof}

The following lemma states that moments of $K(v)^{\gamma}$ exist
for any postive constant $\gamma$.

\begin{lemma} \label{lem:K_alpha} Assume Condition \ref{kernel_prop}.
For any constant $\gamma\in(0,\infty)$ and coordinates $(v_{i_{1}},\cdots,v_{i_{l}})$ of $v$ with $l\leq p+6$,
$\int\prod_{k=1}^{l}v_{i_{k}}K(v)^{\gamma}\,dv<\infty$.
\end{lemma}

\begin{proof}
By Condition \ref{kernel_prop} (iv), for some positive constant $M$
there exists $x_{0}\in(0,\infty)$ such that when $\|v\|>x_{0}$ ,
$K(v)<Me^{-c_{1}\|v\|^{\alpha_{1}}}$. Then consider the integration
in two regions $\{v:\|v\|\leq x_{0}\}$ and $\{v:\|v\|>x_{0}\}$ separately.
In the first region, since $K(v)\leq1$, we have 
\[
\int_{\|v\|\leq x_{0}}\prod_{k=1}^{l}v_{i_{k}}K(v)^{\gamma}\,dv\leq x_{0}^{l}V_{x_{0}},
\]
where $V_{x_{0}}$ is the volume of the $d$-dimension sphere with radius
$x_{0}$, and is finite. In the second region, 
\[
\int_{\|v\|>x_{0}}\prod_{k=1}^{l}v_{i_{k}}K(v)^{\gamma}\,dv\leq M\int_{\|v\|>x_{0}}\|v\|^{l}e^{-c_{1}\gamma\|v\|^{\alpha_{1}}}\,dv.
\]
The right hand side of this is proportional to $\exp\{-c_{1}\gamma x_{0}^{\alpha_{1}/(l+d)}\}$
by integrating in spherical coordinates. %Therefore the lemma holds. 
\end{proof}

\begin{proof}[of Theorem \ref{thm:2}]For (i), since $p_{{\rm acc},\pi}=\varepsilon_{n}^{d}\pi(1)$ and $\pi(1)=\Theta_{p}(a_{n,\varepsilon}^{d-p})$ by Lemmas \ref{marglik_ignored}, \ref{posterior_normal} and \ref{lem:norm_remain},
then $p_{{\rm acc},\pi}=\Theta_{p}(\varepsilon_{n}^{d}a_{n,\varepsilon}^{d-p})$. Together with Lemma \ref{ABCvar}, (i) holds.

For (ii), if we can show that $p_{{\rm acc},q}=\Theta_{p}(\varepsilon_{n}^{d}a_{n,\varepsilon}^{d})$,
then the order of $\Sigma_{{\rm IS},n}$ is obvious from \eqref{eq:expression_Sigma_ABC1}
and the definition of $\Sigma_{{\rm ABC},n}$. Similar to the expansion
of $\pi(1)$ from Lemma \ref{marglik_ignored} and \eqref{norm_approx},
\begin{align*}
p_{{\rm acc},q} & =\varepsilon_{n}^{d}\int\pi_{\rm ABC}(\theta\mid s_{{\rm obs}},\varepsilon_{n})f_{\rm ABC}(s_{{\rm obs}}\mid\theta)\,d\theta\\
 & =\varepsilon_{n}^{d}\left\{ \frac{\int_{B_{\delta}}\pi(\theta)\ftil_{\rm ABC}(s_{{\rm obs}}\mid\theta)^{2}\,d\theta}{\tpi_{B_{\delta}}(1)}+O_{p}(\alpha_{n}^{-1})\right\} \{1+o_{p}(1)\}.
\end{align*}
The integral in the above differs from $\tpi_{B_{\delta}}(1)$ by
the square power of $\ftil_{\rm ABC}(s_{{\rm obs}}\mid\theta)$ in the
integrand. We will show that this integral has order $\Theta_{p}(a_{n,\varepsilon}^{2d-p})$,
from which $p_{{\rm acc},q}=\Theta_{p}(\varepsilon_{n}^{d}a_{n,\varepsilon}^{d})$
trivially holds. Let $g_{n}^{**}(t,v;\tau_{1},\tau_{2})$ be the function
\begin{align*}
g_{n}^{**}(t,v;\tau_{1},\tau_{2}) & =
N[v;0,\{A(\theta_{0})^{-1/2}+\tau_{2}\}^{-2}]
K\left[\{Ds(\theta_{0})+\tau_{1}\}t+\frac{1}{a_{n}\varepsilon_{n}}v^{*}-\frac{1}{a_{n}\varepsilon_{n}}A(\theta_{0})^{1/2}T_{\rm obs}\right],
\end{align*}
and $g_{n}^{**}(t,v;\tau_{1},\tau_{2})=g_{n}^{*}(t,v+a_{n}\varepsilon_{n}\tau_{1}t;\tau_{1},\tau_{2})$.
Here expansions \eqref{eq:fK_expand1} and \eqref{eq:fK_expand2}
of $\ftil_{n}(s_{{\rm obs}}+\varepsilon_{n}v\mid\theta_{0}+a_{n,\varepsilon}^{-1}t)K(v)$
are to be used in the form of 
\begin{align}
\frac{a_{n,\varepsilon}^{d}\vert A(\theta_{0})\vert^{1/2}}{\vert A(\theta_{0}+a_{n,\varepsilon}^{-1}t)\vert^{1/2}}\begin{cases}
\left\{ g_{n}(t,v)+a_{n}^{-1}P_{3}(t,v)g_{n,r}(t,v)\right\} ,\quad \lim_{n\rightarrow\infty}a_{n}\varepsilon_{n}<\infty,\\
\left\{ g_{n}^{*}(t,v^{*})+a_{n}\varepsilon_{n}^{2}P_{3}(t,v^{*})g_{n}^{*}(t,v^{*})\right.\\
\left.+(a_{n}\varepsilon_{n}^{2})^{2}P_{4}(t,v^{**})g_{n,r}^{*}(t,v^{**})\right\} (a_{n}\varepsilon_{n})^{d},\quad \lim_{n\rightarrow\infty}a_{n}\varepsilon_{n}=\infty,
\end{cases}\label{eq:fK_expandAlt}
\end{align}
where $P_{3}(t,v^{*})\in\mathbb{P}_{3,(t,v^{*})}$, $g_{n,r}(t,v)=g_{n}\{t,v;e_{n1}r_{s}(t,\epsilon_{1}),e_{n1}r_{A}(t,\epsilon_{2})\}$,
$g_{n,r}^{*}(t,v^{**})$ is $g_{n}^{**}\{t,v^{**};e_{n2}r_{s}(t,\epsilon_{3}),e_{n2}r_{A}(t,\epsilon_{4})\}$
and $P_{4}(t,v^{**})$ is $P_{4}(t,v^{*})$ with the transformation
$v^{**}=v^{*}-a_{n}\varepsilon_{n}e_{n2}r_{s}(t,\epsilon_{3})t$,
and the expansion of $\pi(\theta)/\vert A(\theta)\vert$ similar to
\eqref{eq:fK_expand3} is to be used. 
By the expression of $P_{4}(t,v^{*})$
in \eqref{eq:fK_expand2}, it can be seen that $P_{4}(t,v^{**})\in\mathbb{P}_{4,(t,v^{**})}$.
Basic inequalities $(a+\varepsilon b)^{2}\leq\varepsilon a^{2}+(\varepsilon+\varepsilon^{2})b^{2}$
and $(a+\varepsilon b+\varepsilon^{2}c)^{2}\leq(\varepsilon+\varepsilon^{2})a^{2}+(\varepsilon+\varepsilon^{2}+\varepsilon^{3})b^{2}+(\varepsilon^{2}+\varepsilon^{3}+\varepsilon^{4})c^{2}$
for any real constants $a$, $b$, $c$ and $\varepsilon$, from the
fact that $2ab\leq a^{2}+b^{2}$, are also to be used. Then by the above
expansions and inequalities, an expansion of the target integral similar
to \eqref{norm_expand3} can be obtained, with the leading term $a_{n,\varepsilon}^{2d-p}\pi(\theta_{0})\int_{t(B_{\delta})}\{\int g_{n}(t,v)\,dv\}^{2}\,dt$
and remainder term with the following upper bound 
\begin{align*}
 & \Big|a_{n,\varepsilon}^{p-2d}\int_{B_{\delta}}\pi(\theta)\ftil_{\rm ABC}(s_{{\rm obs}}\mid\theta)^{2}\,d\theta-\pi(\theta_{0})\int_{t(B_{\delta})}\Big\{\int g_{n}(t,v)\,dv\Big\}^{2}\,dt\Big|\\
\leq & a_{n,\varepsilon}^{-1}\int_{t(B_{\delta})}\vert A(\theta_{0})\vert D_{\theta}\frac{\pi(\theta_{0}+\epsilon_{6})}{\vert A(\theta_{0}+\epsilon_{6})\vert}t\Big\{\int g_{n}(t,v)\,dv\Big\}^{2}\,dt\\
 & +M\int_{t(B_{\delta})}\Big[a_{n}^{-1}\Big\{\int g_{n}(t,v)\,dv\Big\}^{2}+(a_{n}^{-1}+a_{n}^{-2})\Big\{\int P_{3}(t,v)g_{n,r}(t,v)\,dv\Big\}^{2}\Big]\,dt\mathbbm{1}_{\{\lim a_{n}\varepsilon_{n}<\infty\}}\\
 & +M\int_{t(B_{\delta})}\Big[\{a_{n}\varepsilon_{n}^{2}+(a_{n}\varepsilon_{n}^{2})^{2}\}\Big\{\int g_{n}^{*}(t,v^{*})\,dv^{*}\Big\}^{2}\\
 & \ \ \ \ +\{a_{n}\varepsilon_{n}^{2}+(a_{n}\varepsilon_{n}^{2})^{2}+(a_{n}\varepsilon_{n}^{2})^{3}\}\Big\{\int P_{3}(t,v^{*})g_{n}^{*}(t,v^{*})\,dv^{*}\Big\}^{2}\\
 & \ \ \ \ +\{(a_{n}\varepsilon_{n}^{2})^{2}+(a_{n}\varepsilon_{n}^{2})^{3}+(a_{n}\varepsilon_{n}^{2})^{4}\}\Big\{\int P_{4}(t,v^{**})g_{n,r}^{*}(t,v^{**})\,dv^{**}\Big\}^{2}\Big]\,dt\mathbbm{1}_{\{\lim a_{n}\varepsilon_{n}=\infty\}},
\end{align*}
where $M$ is the upper bound of $\pi(\theta)\vert A(\theta_{0})\vert/\vert A(\theta)\vert$
for $\theta\in B_{\delta}$ with $\delta$ chosen so that $M$ exists.
Then if we can show that for any $P_{4}(t,v)\in\mathbb{P}_{5,(t,v)}$,
$d\times p$ matrix function $r_{n1}(t)$ and $d\times d$ matrix
function $r_{n2}(t)$ which can be bounded by $\sigma I_{d}$ and
$\sigma I_{p}$ uniformly in $n$ and $t$ for any small $\delta$
if $\delta$ is small enough, (a)$\int_{t(B_{\delta})}\left\{ \int_{\mathbb{R}^{d}}g_{n}(t,v)\,dv\right\} ^{2}\,dt$
is $\Theta_{p}(1)$; (b) $\int_{t(B_{\delta})}\left[ \int_{\mathbb{R}^{d}}P_{4}(t,v)g_{n}\{t,v;r_{n1}(t),r_{n2}(t)\}\,dv\right] ^{2}\,dt$
is $O_{p}(1)$ when $\lim_{n\rightarrow\infty}a_{n}\varepsilon_{n}<\infty$;
(c) $\int_{t(B_{\delta})}\left[ \int_{\mathbb{R}^{d}}P_{4}(t,v)g_{n}^{**}\{t,v;r_{n1}(t),r_{n2}(t)\}\,dv\right] ^{2}\,dt$
is $O_{p}(1)$ when $\lim_{n\rightarrow\infty}a_{n}\varepsilon_{n}=\infty$,
the lemma would hold. 

Here $\delta$ is selected such that $Ds(\theta_{0})+r_{n1}(t)$ is
bounded bounded by $m_{1}I_{p}$ and $m_{2}I_{d}\leq A(\theta_{0})^{-1/2}+r_{n2}(t)\leq M_{2}I_{d}$,
for some positive constants $m_{1}$, $m_{2}$ and $M_{2}$, uniformly
in $n$ and $t$. For the purpose of bounding integrals, we can assume
that $A(\theta_{0})=I_{d}$ and $r_{n2}(t)=0$ without loss of generality
by the following inequality when $\lim_{n\rightarrow\infty}a_{n}\varepsilon_{n}<\infty$,
\[
g_{n}\{t,v;r_{n1}(t),r_{n2}(t)\}\leq\frac{M_{2}^{d}}{(2\pi)^{d/2}}\exp\Big[-\frac{m_{2}^{2}}{2}\|a_{n}\varepsilon_{n}v+A(\theta_{0})^{1/2}T_{\rm obs}-\{Ds(\theta_{0})+r_{n1}(t)\}t\|^{2}\Big] K(v),
\]
and a similar one for $g_{n}^{**}\{t,v;r_{n1}(t),r_{n2}(t)\}$.

Consider any $P_{4}(t,v)\in\mathbb{P}_{4,(t,v)}$. When $\lim_{n\rightarrow\infty}a_{n}\varepsilon_{n}<\infty$,
let $E_{1}=\{v:\|a_{n}\varepsilon_{n}v\|^{2}\leq\beta_{1}\|\{Ds(\theta_{0})+r_{n1}(t)\}t-A(\theta_{0})^{1/2}T_{{\rm obs}}\|^{2}\}$
for some $\beta_{1}\in(0,1)$. Then for any $\beta_{2}\in(0,1)$ we
have
\begin{align}
 & \int_{\mathbb{R}^{d}}P_{4}(t,v)g_{n}\{t,v;r_{n1}(t),r_{n2}(t)\}\,dv\nonumber \\
\leq & \left(\int_{E_{1}}+\int_{E_{1}^{c}}\right)P_{4}(t,v)\frac{M_{2}^{d}}{(2\pi)^{d/2}}
\exp\left[ -\frac{m_{2}^{2}}{2}\|a_{n}\varepsilon_{n}v-\{Ds(\theta_{0})+r_{n1}(t)\}t+A(\theta_{0})^{1/2}T_{{\rm obs}}\|^{2}\right] K(v)\,dv\nonumber \\
\leq & P_{4}(t)\left(\exp\left[ -\frac{m_{2}^{2}(1-\beta_{1})}{2}\|\{Ds(\theta_{0})+r_{n1}(t)\}t-A(\theta_{0})^{1/2}T_{{\rm obs}}\|^{2}\right] \right.\nonumber \\
 & \left.+\overline{K}^{\beta_{2}}\left[\frac{\lambda_{{\rm min}}^{2}(\Lambda)\beta_{1}}{a_{n}^{2}\varepsilon_{n}^{2}}\|\{Ds(\theta_{0})+r_{n1}(t)\}t-A(\theta_{0})^{1/2}T_{{\rm obs}}\|^{2}\right]\right),\label{eq:integral_v1}
\end{align}
where $P_{4}(t)\in\mathbb{P}_{4,t}$ and the above inequality uses
Lemma \ref{lem:K_alpha}. Then using $(a+b)^{2}\leq2(a^{2}+b^{2})$,
\begin{align*}
 & \int_{t(B_{\delta})}\left[ \int_{\mathbb{R}^{d}}P_{4}(t,v)g_{n}\{t,v;r_{n1}(t),r_{n2}(t)\}\,dv\right]^{2}dt\\
\leq & \int_{t(B_{\delta})}P_{8}(t)\exp\left[ -m_{2}^{2}(1-\beta_{1})\|\{Ds(\theta_{0})+r_{n1}(t)\}t-A(\theta_{0})^{1/2}T_{{\rm obs}}\|^{2}\right] \,dt\\
 & +\int_{t(B_{\delta})}P_{8}(t)\overline{K}^{2\beta_{2}}
 \left[\frac{\lambda_{{\rm min}}^{2}(\Lambda)\beta_{1}}{a_{n}^{2}\varepsilon_{n}^{2}}\|\{Ds(\theta_{0})+r_{n1}(t)\}t-A(\theta_{0})^{1/2}T_{{\rm obs}}\|^{2}\right]\,dt,
\end{align*}
where $P_{8}(t)\in\mathbb{P}_{8,t}$. 

When $a_{n}\varepsilon_{n}\rightarrow\infty$, let $E_{2}=\{v:\|(a_{n}\varepsilon_{n})^{-1}v\|^{2}\leq\beta_{1}\|\{Ds(\theta_{0})+r_{n1}(t)\}t-(a_{n}\varepsilon_{n})^{-1}A(\theta_{0})^{1/2}T_{{\rm obs}}\|^{2}\}$
for some $\beta_{1}\in(0,1)$. Then for any $\beta_{2}\in(0,1)$ we
have 
\begin{align}
 & \int_{\mathbb{R}^{d}}P_{4}(t,v)g_{n}^{**}\{t,v;r_{n1}(t),r_{n2}(t)\}\,dv\nonumber \\
\leq & \left(\int_{E_{2}}+\int_{E_{2}^{c}}\right)P_{4}(t,v)
K\left[\frac{1}{a_{n}\varepsilon_{n}}v+\{Ds(\theta_{0})+r_{n1}(t)\}t-\frac{1}{a_{n}\varepsilon_{n}}A(\theta_{0})^{1/2}T_{\rm obs}\right] \\ & \times \frac{M_{2}^{d}}{(2\pi)^{d/2}}
\exp\left( -\frac{m_{2}^{2}}{2}\|v\|^{2}\right) \,dv\nonumber \\
\leq & P_{4}(t)\left(\overline{K}\left[\lambda_{{\rm min}}^{2}(\Lambda)(1-\beta_{1})\|\{Ds(\theta_{0})+r_{n1}(t)\}t-\frac{1}{a_{n}\varepsilon_{n}}A(\theta_{0})^{1/2}T_{{\rm obs}}\|^{2}\right]\right.\nonumber \\
 & \left.+\exp\left[ -\frac{a_{n}^{2}\varepsilon_{n}^{2}\beta_{1}m_{2}^{2}\beta_{2}}{2}\|\{Ds(\theta_{0})+r_{n1}(t)\}t-\frac{1}{a_{n}\varepsilon_{n}}A(\theta_{0})^{1/2}T_{{\rm obs}}\|^{2}\right] \right),\label{eq:integral_v2}
\end{align}
where $P_{4}(t)\in\mathbb{P}_{4,t}$. Then using $(a+b)^{2}\leq2(a^{2}+b^{2})$, 
\begin{align*}
 & \int_{t(B_{\delta})}\left[\int_{\mathbb{R}^{d}}P_{4}(t,v)g_{n}^{**}\{t,v;r_{n1}(t),r_{n2}(t)\}\,dv\right]^{2}dt\\
\leq & \int_{t(B_{\delta})}P_{8}(t)
\overline{K}^{2}\left[\frac{\lambda_{{\rm min}}^{2}(\Lambda)(1-\beta_{1})}{2}\|\{Ds(\theta_{0})+r_{n1}(t)\}t-\frac{1}{a_{n}\varepsilon_{n}}A(\theta_{0})^{1/2}T_{{\rm obs}}\|^{2}\right]\,dt\\
 & +\int_{t(B_{\delta})}P_{8}(t)\exp\left[ 
 -\frac{a_{n}^{2}\varepsilon_{n}^{2}\beta_{1}m_{2}^{2}\beta_{2}}{2}\|\{Ds(\theta_{0})+r_{n1}(t)\}t-\frac{1}{a_{n}\varepsilon_{n}}A(\theta_{0})^{1/2}T_{{\rm obs}}\|^{2}\right] \,dt
\end{align*}
Applying Lemma \ref{poly_mixnorm_bound} on these upper bounds,
(b) and (c) hold.

For (a), to see that the limit of $\int_{t(B_{\delta})}\{\int_{\mathbb{R}^{d}}g_{n}(t,v)\,dv\}^{2}\,dt$
is lower bounded away from zero, just use the positivity of the limit
of the integrand and Fatou's lemma to interchange the order of limit
and integral. \end{proof}

\subsection{Proof of Theorem 3}

Now let $w_{n}(\theta)$ be the importance weight $\pi(\theta)/q_{n}(\theta)$,
define $\pi_{B_{\delta},{\rm IS}}(h)=\int_{B_{\delta}}h(\theta)\pi(\theta)f_{\rm ABC}(s_{{\rm obs}}\mid\theta)w_{n}(\theta)\ d\theta$
and define $\pi_{B_{\delta}^{c},{\rm IS}}(h)$ correspondingly. Then by
\eqref{eq:expression_Sigma_ABC1}, we have 
\begin{align}
\Sigma_{ABC,n} & =p_{{\rm acc},\pi}^{-1}\frac{\pi_{B_{\delta},{\rm IS}}(G_{n})+\pi_{B_{\delta}^{c},{\rm IS}}(G_{n})}{\pi_{B_{\delta}}(1)+\pi_{B_{\delta}^{c}}(1)}.\label{eq:expression_Sigma_ABC2}
\end{align}

\begin{proof}[ of Theorem \ref{thm:3}] For $p_{{\rm acc},q_{n}}$,
we only need to consider the case when $\beta=0$. Recall that $t(\theta)=a_{n,\varepsilon}(\theta-\theta_{0})$.
By the transformation $t=t(\theta)$, since $a_{n,\varepsilon}\sigma_{n}=1$,
$q_{n}(\theta)=a_{n,\varepsilon}^{p}|\Sigma|^{-1/2}q\{\Sigma^{-1/2}(t-c_{\mu})\}$.
Then, similar to the expansion of $\pi(1)$ from Lemma \ref{marglik_ignored},
%and \eqref{norm_approx}, 
\begin{align*}
p_{{\rm acc},q_{n}} & =\varepsilon_{n}^{d}\int q_{n}(\theta)f_{\rm ABC}(s_{{\rm obs}}\mid\theta)\,d\theta\\
 & =\varepsilon_{n}^{d}|\Sigma|^{-1/2}\int_{t(B_{\delta})}q\{\Sigma^{-1/2}(t-c_{\mu})\}\ftil_{\rm ABC}(s_{{\rm obs}}\mid\theta_{0}+a_{n,\varepsilon}^{-1}t)\,dt\{1+o_{p}(1)\}.
\end{align*}
The above integral differs from $\widetilde{\pi}_{B_{\delta}}(1)$
by replacing $\pi(\theta_{0}+a_{n,\varepsilon}^{-1}t)$ with the density
$q\{\Sigma^{-1/2}(t-c_{\mu})\}$ which does not degenerate to a constant
as $n\rightarrow\infty$. We will show that this integral has order
$\Theta_{p}(1)$. Plugging in the expansion \eqref{eq:fK_expandAlt}
of $\ftil_{\rm ABC}(s_{{\rm obs}}\mid\theta_{0}+a_{n,\varepsilon}^{-1}t)$
into $p_{{\rm acc},q_{n}}$, we can obtain an expansion similar to \eqref{norm_expand3},
differing in that parts from expanding $\pi(\theta_{0}+a_{n,\varepsilon}^{-1}t)/\vert A(\theta_{0}+a_{n,\varepsilon}^{-1}t)\vert^{1/2}$
are replaced by the Taylor expansion 
\[
\frac{q\{\Sigma^{-1/2}(t-c_{\mu})\}}{\vert A(\theta_{0}+a_{n,\varepsilon}^{-1}t)\vert^{1/2}}=q\{\Sigma^{-1/2}(t-c_{\mu})\}\left[ 
1+a_{n,\varepsilon}^{-1}D_{\theta}\frac{1}{\vert A\{\theta_{0}+\epsilon_{6}(t)\}\vert^{1/2}}t\right] ,
\]
where $\|\epsilon_{6}(t)\|\leq\delta$. The explicit form is ommitted
here to avoid repetition. It can be seen that $p_{{\rm acc},q_{n}}=\Theta_{p}(a_{n,\varepsilon}^{d}\varepsilon_{n}^{d})$
if (a) $\int_{\mathbb{R}^{d}\times t(B_{\delta})}q\{\Sigma^{-1/2}(t-c_{\mu})\}g_{n}(t,v)\,dvdt=\Theta_{p}(1)$;
(b) $\int_{\mathbb{R}^{d}\times t(B_{\delta})}P_{3}(t,v)q\{\Sigma^{-1/2}(t-c_{\mu})\}g_{n}\{t,v;r_{n1}(t),r_{n2}(t)\}\,dvdt=O_{p}(1)$
when $\lim_{n\rightarrow\infty}a_{n}\varepsilon_{n}<\infty$; and (c)
$\int_{\mathbb{R}^{d}\times t(B_{\delta})}P_{3}(t,v)q\{\Sigma^{-1/2}(t-c_{\mu})\}g_{n}^{**}\{t,v;r_{n1}(t),r_{n2}(t)\}\,dvdt=O_{p}(1)$
when $\lim_{n\rightarrow\infty}a_{n}\varepsilon_{n}=\infty$, where
$r_{n1}(t)$ and $r_{n2}(t)$ are defined as in the proof of Theorem
\ref{thm:2}. Since $q\{\Sigma^{-1/2}(t-c_{\mu})\}$ is uniformly
upper bounded for $t\in\mathbb{R}^{p}$, (b) and (c) hold and the
integral in (a) is $O_{p}(1)$ following the arguments for the similar
cases in the proof of Theorem \ref{thm:2}. By the positivity
of the limit of the integrand and Fatou's lemma, the limit of the
integral in (a) is lower bounded away from $0$. Therefore $p_{{\rm acc},q_{n}}=\Theta_{p}(a_{n,\varepsilon}^{d}\varepsilon_{n}^{d})$
holds.

As $\Sigma_{{\rm IS},n}$ is equal to $p_{{\rm acc},q_{n}}\Sigma_{ABC,n}$, by \eqref{eq:expression_Sigma_ABC1} we have 
\begin{align*}
\Sigma_{{\rm IS},n} & =\frac{p_{{\rm acc},q_{n}}}{p_{{\rm acc},\pi}}\frac{\pi_{B_{\delta},{\rm IS}}(G_{n})+\pi_{B_{\delta}^{c},{\rm IS}}(G_{n})}{\pi_{B_{\delta}}(1)+\pi_{B_{\delta}^{c}}(1)}=\frac{p_{{\rm acc},q_{n}}}{p_{{\rm acc},\pi}}\frac{\pi_{B_{\delta},{\rm IS}}(G_{n})}{\pi_{B_{\delta}}(1)}\{1+o_{p}(1)\},
\end{align*}
where the second equality holds by noting that $\omega_{n}(\theta)\leq\beta^{-1}$.
Given the obtained orders of $p_{{\rm acc},q_{n}}$ and $p_{{\rm acc},\pi}$,
 $\Sigma_{{\rm IS},n}=O_{p}(a_{n,\varepsilon}^{-2})$ if $\pi_{B_{\delta},{\rm IS}}(G_{n})/\pi_{B_{\delta}}(1)=O_{p}(a_{n,\varepsilon}^{-p-2})$.
Similar to \eqref{eq:expansion_variance}, we have the following expansion
\begin{eqnarray*}
\lefteqn{\frac{\pi_{B_{\delta},{\rm IS}}(G_{n})}{\pi_{B_{\delta}}(1)} =G(\theta_{0})\frac{\pi_{B_{\delta},{\rm IS}}(1)}{\pi_{B_{\delta}}(1)}}\\
&+& 2a_{n,\varepsilon}^{-1}\{h(\theta_{0})-h_{\rm ABC}\}\frac{\pi_{B_{\delta},{\rm IS}}\{Dh(\theta_{t})^{T}t\}}{\pi_{B_{\delta}}(1)}
+a_{n,\varepsilon}^{-2}\frac{\pi_{B_{\delta},{\rm IS}}\{t^{T}Dh(\theta_{t})Dh(\theta_{t})^{T}t\}}{\pi_{B_{\delta}}(1)},
\end{eqnarray*}
and we only need $\pi_{B_{\delta},{\rm IS}}\{P_{2}(t)\}/\pi_{B_{\delta}}(1)=O_{p}(a_{n,\varepsilon}^{-p})$
for any $P_{2}(t)\in\mathbb{P}_{2,t}$. Since $w_{n}(\theta)\leq(1-\beta)^{-1}w_{n,0}(\theta)$,
where $w_{n,0}(\theta)$ is the weight when $\beta=0$, it is sufficient
to consider the case $\beta=0$. 

Similar to the proof of Theorem \ref{thm:1}, first the normal counterpart
$\widetilde{\pi}_{B_{\delta},{\rm IS}}\{P_{2}(t)\}/\widetilde{\pi}_{B_{\delta}}(1)$
of $\pi_{B_{\delta},{\rm IS}}\{P_{2}(t)\}/\pi_{B_{\delta}}(1)$, where $f_{\rm ABC}(s_{{\rm obs}}\mid\theta)$
is replaced by $\ftil_{\rm ABC}(s_{{\rm obs}}\mid\theta)$, is considered,
then it is shown that their difference can be ignored. Using the transformation
$t=t(\theta)$ and plugging in expansion \eqref{eq:fK_expandAlt}
of $\ftil_{\rm ABC}(s_{{\rm obs}}\mid\theta_{0}+a_{n,\varepsilon}^{-1}t)$
into $\widetilde{\pi}_{B_{\delta},{\rm IS}}\{P_{2}(t)\}$, we obtain an
expansion similar to \eqref{norm_expand3}, differing in that parts
from expanding $\pi(\theta_{0}+a_{n,\varepsilon}^{-1}t)/\vert A(\theta_{0}+a_{n,\varepsilon}^{-1}t)\vert^{1/2}$
are replaced by the Taylor expansion 
\begin{eqnarray*}
\lefteqn{\frac{1}{q_{n}(\theta)}\frac{\pi(\theta_{0}+a_{n,\varepsilon}^{-1}t)^{2}}{\vert A(\theta_{0}+a_{n,\varepsilon}^{-1}t)\vert^{1/2}} }\\
&=&\frac{1}{a_{n,\varepsilon}^{p}\vert\Sigma\vert^{-1/2}q\{\Sigma^{-1/2}(t-c_{\mu})\}}
\left[ \pi(\theta_{0})^{2}+a_{n,\varepsilon}^{-1}D_{\theta}\frac{\pi\{\theta_{0}+\epsilon_{7}(t)\}^{2}}{\vert A\{\theta_{0}+\epsilon_{7}(t)\}\vert^{1/2}}t\right] ,
\end{eqnarray*}
where $\|\epsilon_{7}(t)\|\leq\delta$. The explicit form is omitted
here to avoid repetition. Then it can be seen that if we can show
that 
\begin{align*}
(d)\ \int_{t(B_{\delta})}\frac{\int_{\mathbb{R}^{d}}P_{5}(t,v)g_{n}\{t,v;r_{n1}(t),r_{n2}(t)\}\,dv}{q\{\Sigma^{-1/2}(t-c_{\mu})\}}\,dt=O_{p}(1) & \text{ when }\lim_{n\rightarrow\infty}a_{n}\varepsilon_{n}<\infty,\\
(e)\ \int_{t(B_{\delta})}\frac{\int_{\mathbb{R}^{d}}P_{5}(t,v)g_{n}^{**}\{t,v;r_{n1}(t),r_{n2}(t)\}\,dv}{q\{\Sigma^{-1/2}(t-c_{\mu})\}}\,dt=O_{p}(1) & \text{ when }\lim_{n\rightarrow\infty}a_{n}\varepsilon_{n}=\infty,
\end{align*}
where $r_{n1}(t)$ and $r_{n2}(t)$ are defined as in the proof of
Theorem \ref{thm:2}, $\widetilde{\pi}_{B_{\delta},{\rm IS}}\{P_{2}(t)\}=O_{p}(a_{n,\varepsilon}^{d-2p})$
and $\widetilde{\pi}_{B_{\delta},{\rm IS}}\{P_{2}(t)\}/\widetilde{\pi}_{B_{\delta}}(1)=O_{p}(a_{n,\varepsilon}^{-p})$
by Lemma \ref{posterior_normal}. By \eqref{eq:integral_v1} and the
following equality for $d\times p$ full column-rank matrix $A$ and
vector $c$,
\[
\|At-c\|=\|P^{1/2}(t-P^{-1}Ac)\|^{2}+c^{T}(I-AP^{-1}A^{T})c,
\]
where $P=A^{T}A$ and $P^{1/2}P^{1/2}=P$, for (d) we have 
\begin{align*}
 & \frac{\int_{\mathbb{R}^{d}}P_{5}(t,v)g_{n}\{t,v;r_{n1}(t),r_{n2}(t)\}\,dv}{q\{\Sigma^{-1/2}(t-c_{\mu})\}}\\
\leq & P_{5}(t)\frac{\exp\left\{ -\frac{m_{1}^{2}m_{2}^{2}\gamma}{2}\|t-P(\theta_{0},t)T_{\rm obs}\|^{2}\right\} }{q\{\Sigma^{-1/2}(t-c_{\mu})\}}
\exp\left[ -\frac{m_{2}^{2}\Delta}{2}\|\{Ds(\theta_{0})+r_{n1}(t)\}t-A(\theta_{0})^{1/2}T_{\rm obs}\|^{2}\right] \\
 & +P_{5}(t)\frac{\overline{K}^{\alpha}\left\{\frac{\lambda_{{\rm min}}^{2}(\Lambda)(1-\gamma-\Delta)m_{1}^{2}}{a_{n}^{2}\varepsilon_{n}^{2}}\|t-P(\theta_{0},t)T_{\rm obs}\|^{2}\right\}}
 {q\{\Sigma^{-1/2}(t-c_{\mu})\}}\\
 & \times \overline{K}^{\Delta}\left[
 \frac{\lambda_{{\rm min}}^{2}(\Lambda)(1-\gamma-\Delta)}{a_{n}^{2}\varepsilon_{n}^{2}}\|\{Ds(\theta_{0})+r_{n1}(t)\}t-A(\theta_{0})^{1/2}T_{\rm obs}\|^{2}\right],
\end{align*}
where $P(\theta_{0},t)=[\{Ds(\theta_{0})+r_{n1}(t)\}^{T}\{Ds(\theta_{0})+r_{n1}(t)\}]^{-1}\{Ds(\theta_{0})+r_{n1}(t)\}^{T}A(\theta_{0})^{1/2}$,
both $P_{5}(t)$ belong to $\mathbb{P}_{5,t}$ and $\Delta$ is chosen
such that $\gamma+\Delta\in(0,1)$ and $\alpha+\Delta\in(0,1)$ for
$\gamma$ and $\alpha$ in Condition \ref{approp_proposal}. Then
since both ratios on the right hand side of the above inequality are
$O_{p}(1)$ by Condition \ref{approp_proposal}, by Lemma \ref{poly_mixnorm_bound}
and Lemma \ref{lem:K_alpha}, (d) holds. Similarly
by \eqref{eq:integral_v2}, for (e) we have 
\begin{align*}
 & \frac{\int_{\mathbb{R}^{d}}P_{5}(t,v)g_{n}^{**}\{t,v;r_{n1}(t),r_{n2}(t)\}\,dv}{q\{\Sigma^{-1/2}(t-c_{\mu})\}}\\
\leq & P_{5}(t)\frac{\exp\left\{ -\frac{m_{1}^{2}m_{2}^{2}\gamma}{2}\|t-\frac{1}{a_{n}\varepsilon_{n}}P(\theta_{0},t)T_{\rm obs}\|^{2}\right\} }{q\{\Sigma^{-1/2}(t-c_{\mu})\}} \\
& \times
\exp\left[ -\frac{(a_{n}^{2}\varepsilon_{n}^{2}\beta_{1}\beta_{2}-\gamma)m_{2}^{2}}{2}\|\{Ds(\theta_{0})+r_{n1}(t)\}t-A(\theta_{0})^{1/2}T_{\rm obs}\|^{2}\right] \\
 & +P_{5}(t)\frac{\overline{K}^{\alpha}\left\{\lambda_{{\rm min}}^{2}(\Lambda)(1-\beta_{1})m_{1}^{2}\|t-\frac{1}{a_{n}\varepsilon_{n}}P(\theta_{0},t)T_{\rm obs}\|^{2}\right\}}
 {q\{\Sigma^{-1/2}(t-c_{\mu})\}}\\
 &\times \overline{K}^{1-\alpha}\left[\lambda_{{\rm min}}^{2}(\Lambda)(1-\beta_{1})\|\{Ds(\theta_{0})+r_{n1}(t)\}t-A(\theta_{0})^{1/2}T_{\rm obs}\|^{2}\right],
\end{align*}
where both $P_{5}(t)$ belong to $\mathbb{P}_{5,t}$. Thus by Condition
\ref{approp_proposal}, Lemma \ref{poly_mixnorm_bound} and Lemma
\ref{lem:K_alpha}, (e) holds. Therefore $\widetilde{\pi}_{B_{\delta},{\rm IS}}\{P_{2}(t)\}/\widetilde{\pi}_{B_{\delta}}(1)=O_{p}(a_{n,\varepsilon}^{-p})$.

To show that $\pi_{B_{\delta},{\rm IS}}\{P_{2}(t)\}/\pi_{B_{\delta}}(1)=O_{p}(a_{n,\varepsilon}^{-p})$,
similar to the discussion of \eqref{norm_approx}, it is sufficient to
show that 
\begin{equation}
\frac{\pi_{B_{\delta},{\rm IS}}\{P_{2}(t)\}-\widetilde{\pi}_{B_{\delta},{\rm IS}}\{P_{2}(t)\}}{\widetilde{\pi}_{B_{\delta}}(1)}=O_{p}(\alpha_{n}^{-1}a_{n,\varepsilon}^{-p}).\label{eq:IS_remain}
\end{equation}
With the transformation $t=t(\theta)$ we have $\pi_{B_{\delta},{\rm IS}}\{P_{2}(t)\}-\widetilde{\pi}_{B_{\delta},{\rm IS}}\{P_{2}(t)\}$ is equal to
\begin{align*}
%\pi_{B_{\delta},{\rm IS}}(P_{2}(t))-\widetilde{\pi}_{B_{\delta},{\rm IS}}(P_{2}(t)) & =
\alpha_{n}^{-1}a_{n,\varepsilon}^{-2p}\int_{t(B_{\delta})}\int P_{2}(t)\pi(\theta_{0}+a_{n,\varepsilon}^{-1}t)^{2}\frac{r_{n}(s_{{\rm obs}}+\varepsilon_{n}v\mid\theta_{0}+a_{n,\varepsilon}^{-1}t)K(v)}{\vert\Sigma\vert^{-1/2}q\{\Sigma^{-1/2}(t-c_{\mu})\}}\,dvdt.
\end{align*}
Then by following the arguments of the proof of Lemma \ref{lem:norm_remain},
we have 
\begin{align*}
 & |\pi_{B_{\delta},{\rm IS}}\{P_{2}(t)\}-\widetilde{\pi}_{B_{\delta},{\rm IS}}\{P_{2}(t)\}|\leq\alpha_{n}^{-1}a_{n,\varepsilon}^{d-2p}\sup_{\theta\in B_{\delta}}\vert\pi(\theta)^{2}A(\theta)^{-1/2}\vert\\
 & \times \int_{t(B_{\delta})}\int|P_{2}(t)|\frac{(a_{n}a_{n,\varepsilon}^{-1})^{d}r_{{\rm max}}\Big[
 a_{n}a_{n,\varepsilon}^{-1}M\Big\{ Ds(\theta_{0}+\epsilon_t)t-a_{n,\varepsilon}\varepsilon_{n}v-\frac{1}{a_{n}a_{n,\varepsilon}^{-1}}A(\theta_{0})^{1/2}T_{\rm obs}\Big\}
 \Big]K(v)}{q\{\Sigma^{-1/2}(t-c_{\mu})\}}\,dvdt.
\end{align*}
The ratio above is similar to the ratio of $g_{n}\{t,v;r_{1}(t),r_{2}(t)\}/q\{\Sigma^{-1/2}(t-c_{\mu})\}$
except that the normal density is replaced by $r_{{\rm max}}(\cdot)$.
Then by Condition \ref{approp_proposal}, previous arguments for proving
(iv) and (v) can be followed. Hence $\pi_{B_{\delta},{\rm IS}}\{P_{2}(t)\}-\widetilde{\pi}_{B_{\delta},{\rm IS}}\{P_{2}(t)\}=O_{p}(\alpha_{n}a_{n,\varepsilon}^{d-2p})$
and \eqref{eq:IS_remain} holds. Therefore $\Sigma_{{\rm IS},n}=O_{p}(a_{n,\varepsilon}^{-2})$.
\end{proof} 

\bibliographystyle{biometrika}
\bibliography{reference}

\begin{thebibliography}{31}
\expandafter\ifx\csname natexlab\endcsname\relax\def\natexlab#1{#1}\fi

\bibitem[{Allingham et~al.({2009})Allingham, King \&
  Mengersen}]{Allingham:2008}
\textsc{Allingham, D.}, \textsc{King, R. A.~R.} \& \textsc{Mengersen, K.~L.}
  ({2009}).
\newblock {Bayesian estimation of quantile distributions}.
\newblock \textit{{Statistics and Computing}} \textbf{{19}}, {189--201}.

\bibitem[{Barber et~al.(2015)Barber, Voss, Webster et~al.}]{barber2015rate}
\textsc{Barber, S.}, \textsc{Voss, J.}, \textsc{Webster, M.} et~al. (2015).
\newblock The rate of convergence for approximate {B}ayesian computation.
\newblock \textit{Electronic Journal of Statistics} \textbf{9}, 80--105.

\bibitem[{Beaumont(2010)}]{beaumont2010approximate}
\textsc{Beaumont, M.~A.} (2010).
\newblock Approximate {B}ayesian computation in evolution and ecology.
\newblock \textit{Annual Review of Ecology, Evolution, and Systematics}
  \textbf{41}, 379--406.

\bibitem[{Beaumont et~al.(2009)Beaumont, Cornuet, Marin \&
  Robert}]{beaumont2009adaptive}
\textsc{Beaumont, M.~A.}, \textsc{Cornuet, J.-M.}, \textsc{Marin, J.-M.} \&
  \textsc{Robert, C.~P.} (2009).
\newblock Adaptive approximate {B}ayesian computation.
\newblock \textit{Biometrika} \textbf{96}, 983--990.

\bibitem[{Beaumont et~al.(2002)Beaumont, Zhang \& Balding}]{Beaumont:2002}
\textsc{Beaumont, M.~A.}, \textsc{Zhang, W.} \& \textsc{Balding, D.~J.} (2002).
\newblock Approximate {B}ayesian computation in population genetics.
\newblock \textit{Genetics} \textbf{162}, 2025--2035.

\bibitem[{Biau et~al.(2015)Biau, C{\'e}rou \& Guyader}]{biau2015new}
\textsc{Biau, G.}, \textsc{C{\'e}rou, F.} \& \textsc{Guyader, A.} (2015).
\newblock New insights into approximate {B}ayesian computation.
\newblock \textit{Annales de l'Institut Henri Poincar{\'e}, Probabilit{\'e}s et
  Statistiques} \textbf{51}, 376--403.

\bibitem[{Blum(2010)}]{blum2010approximate}
\textsc{Blum, M.~G.} (2010).
\newblock Approximate {B}ayesian computation: a nonparametric perspective.
\newblock \textit{Journal of the American Statistical Association}
  \textbf{105}, 1178--1187.

\bibitem[{Blum \& Fran{\c{c}}ois(2010)}]{blum2010non}
\textsc{Blum, M.~G.} \& \textsc{Fran{\c{c}}ois, O.} (2010).
\newblock Non-linear regression models for approximate {B}ayesian computation.
\newblock \textit{Statistics and Computing} \textbf{20}, 63--73.

\bibitem[{Bortot et~al.(2007)Bortot, Coles \& Sisson}]{bortot2007inference}
\textsc{Bortot, P.}, \textsc{Coles, S.~G.} \& \textsc{Sisson, S.~A.} (2007).
\newblock Inference for stereological extremes.
\newblock \textit{Journal of the American Statistical Association}
  \textbf{102}, 84--92.

\bibitem[{Clarke \& Ghosh(1995)}]{clarke1995posterior}
\textsc{Clarke, B.} \& \textsc{Ghosh, J.} (1995).
\newblock Posterior convergence given the mean.
\newblock \textit{Annals of Statistics} \textbf{23}, 2116--2144.

\bibitem[{Creel \& Kristensen(2013)}]{creel2013indirect}
\textsc{Creel, M.} \& \textsc{Kristensen, D.} (2013).
\newblock Indirect likelihood inference (revised).
\newblock U{F}{A}{E} and {I}{A}{E} working papers, Unitat de Fonaments de
  l'Analisi Economica (UAB) and Institut d'Analisi Economica (CSIC).

\bibitem[{Del~Moral et~al.(2012)Del~Moral, Doucet \& Jasra}]{DelMoral:2012}
\textsc{Del~Moral, P.}, \textsc{Doucet, A.} \& \textsc{Jasra, A.} (2012).
\newblock An adaptive sequential {M}onte {C}arlo method for approximate
  {B}ayesian computation.
\newblock \textit{Statistics and Computing} \textbf{22}, 1009--1020.

\bibitem[{Duffie \& Singleton(1993)}]{duffie1993simulated}
\textsc{Duffie, D.} \& \textsc{Singleton, K.~J.} (1993).
\newblock Simulated moments estimation of {M}arkov models of asset prices.
\newblock \textit{Econometrica} \textbf{61}, 929--952.

\bibitem[{Fearnhead \& Prangle(2012)}]{fearnhead2012constructing}
\textsc{Fearnhead, P.} \& \textsc{Prangle, D.} (2012).
\newblock Constructing summary statistics for approximate {B}ayesian
  computation: semi-automatic approximate {B}ayesian computation (with
  discussion).
\newblock \textit{Journal of the Royal Statistical Society: Series B
  (Statistical Methodology)} \textbf{74}, 419--474.

\bibitem[{Frazier et~al.(2016)Frazier, Martin, Robert \&
  Rousseau}]{Frazier:2017}
\textsc{Frazier, D.~T.}, \textsc{Martin, G.~M.}, \textsc{Robert, C.~P.} \&
  \textsc{Rousseau, J.} (2016).
\newblock {Asymptotic properties of approximate Bayesian computation}.
\newblock \textit{arXiv:1607.06903} .

\bibitem[{Gordon et~al.(1993)Gordon, Salmond \&
  Smith}]{Gordon/Salmond/Smith:1993}
\textsc{Gordon, N.}, \textsc{Salmond, D.} \& \textsc{Smith, A. F.~M.} (1993).
\newblock {Novel approach to nonlinear/non-Gaussian {B}ayesian state
  estimation}.
\newblock \textit{IEEE proceedings F - Radar and Signal Processing}
  \textbf{140}, 107--113.

\bibitem[{Gouri\'{e}roux \& Ronchetti(1993)}]{Gourieroux:1993}
\textsc{Gouri\'{e}roux, C.} \& \textsc{Ronchetti, E.} (1993).
\newblock Indirect inference.
\newblock \textit{Journal of Applied Econometrics} \textbf{8}, s85--s118.

\bibitem[{Heggland \& Frigessi(2004)}]{Frigessi:2004}
\textsc{Heggland, K.} \& \textsc{Frigessi, A.} (2004).
\newblock Estimating functions in indirect inference.
\newblock \textit{Journal of the Royal Statistical Society: Series B
  (Statistical Methodology)} \textbf{66}, 447--462.

\bibitem[{Hesterberg(1995)}]{Hesterberg:1995}
\textsc{Hesterberg, T.} (1995).
\newblock Weighted average importance sampling and defensive mixture
  distributions.
\newblock \textit{Technometrics} \textbf{37}, 185--194.

\bibitem[{Ishida et~al.(2015)Ishida, Vitenti, Penna-Lima, Cisewski, de~Souza,
  Trindade, Cameron \& Busti}]{ishida2015cosmoabc}
\textsc{Ishida, E.}, \textsc{Vitenti, S.}, \textsc{Penna-Lima, M.},
  \textsc{Cisewski, J.}, \textsc{de~Souza, R.}, \textsc{Trindade, A.},
  \textsc{Cameron, E.} \& \textsc{Busti, V.} (2015).
\newblock {COSMOABC}: Likelihood-free inference via population {M}onte {C}arlo
  approximate {B}ayesian computation.
\newblock \textit{Astronomy and Computing} \textbf{13}, 1--11.

\bibitem[{Lehmann(2004)}]{lehmann2004elements}
\textsc{Lehmann, E.~L.} (2004).
\newblock \textit{Elements of large-sample theory}.
\newblock Springer Science \& Business Media.

\bibitem[{Li \& Fearnhead(2018)}]{Li/Fearnhead:2017}
\textsc{Li, W.} \& \textsc{Fearnhead, P.} (2018).
\newblock Convergence of regression adjusted approximate {B}ayesian
  computation.
\newblock \textit{Biometrika} , to appear.

\bibitem[{Marin et~al.(2014)Marin, Pillai, Robert \&
  Rousseau}]{marin2014relevant}
\textsc{Marin, J.-M.}, \textsc{Pillai, N.~S.}, \textsc{Robert, C.~P.} \&
  \textsc{Rousseau, J.} (2014).
\newblock Relevant statistics for {B}ayesian model choice.
\newblock \textit{Journal of the Royal Statistical Society: Series B
  (Statistical Methodology)} \textbf{76}, 833--859.

\bibitem[{Peters et~al.(2011)Peters, Kannan, Lasscock, Mellen, Godsill
  et~al.}]{peters2011bayesian}
\textsc{Peters, G.~W.}, \textsc{Kannan, B.}, \textsc{Lasscock, B.},
  \textsc{Mellen, C.}, \textsc{Godsill, S.} et~al. (2011).
\newblock {B}ayesian cointegrated vector autoregression models incorporating
  alpha-stable noise for inter-day price movements via approximate {B}ayesian
  computation.
\newblock \textit{{B}ayesian Analysis} \textbf{6}, 755--792.

\bibitem[{Prangle et~al.(2014)Prangle, Fearnhead, Cox, Biggs \&
  French}]{prangle2014semi}
\textsc{Prangle, D.}, \textsc{Fearnhead, P.}, \textsc{Cox, M.~P.},
  \textsc{Biggs, P.~J.} \& \textsc{French, N.~P.} (2014).
\newblock Semi-automatic selection of summary statistics for {ABC} model
  choice.
\newblock \textit{Statistical Applications in Genetics and Molecular Biology}
  \textbf{13}, 67--82.

\bibitem[{Pritchard et~al.(1999)Pritchard, Seielstad, {Perez-Lezaun} \&
  Feldman}]{Pritchard:1999}
\textsc{Pritchard, J.~K.}, \textsc{Seielstad, M.~T.}, \textsc{{Perez-Lezaun},
  A.} \& \textsc{Feldman, M.~W.} (1999).
\newblock Population growth of human {Y} chromosomes: a study of {Y} chromosome
  microsatellites.
\newblock \textit{Molecular Biology and Evolution} \textbf{16}, 1791--1798.

\bibitem[{Sandmann \& Koopman(1998)}]{sandmann1998estimation}
\textsc{Sandmann, G.} \& \textsc{Koopman, S.} (1998).
\newblock Estimation of stochastic volatility models via {M}onte {C}arlo
  maximum likelihood.
\newblock \textit{Journal of Econometrics} \textbf{87}, 271--301.

\bibitem[{Toni et~al.(2009)Toni, Welch, Strelkowa, Ipsen \&
  Stumpf}]{toni2009approximate}
\textsc{Toni, T.}, \textsc{Welch, D.}, \textsc{Strelkowa, N.}, \textsc{Ipsen,
  A.} \& \textsc{Stumpf, M.~P.} (2009).
\newblock Approximate {B}ayesian computation scheme for parameter inference and
  model selection in dynamical systems.
\newblock \textit{Journal of the Royal Society Interface} \textbf{6}, 187--202.

\bibitem[{Wegmann et~al.(2009)Wegmann, Leuenberger \&
  Excoffier}]{wegmann2009efficient}
\textsc{Wegmann, D.}, \textsc{Leuenberger, C.} \& \textsc{Excoffier, L.}
  (2009).
\newblock Efficient approximate {B}ayesian computation coupled with {M}arkov
  chain {M}onte {C}arlo without likelihood.
\newblock \textit{Genetics} \textbf{182}, 1207--1218.

\bibitem[{Wood(2010)}]{wood2010statistical}
\textsc{Wood, S.~N.} (2010).
\newblock Statistical inference for noisy nonlinear ecological dynamic systems.
\newblock \textit{Nature} \textbf{466}, 1102--1104.

\bibitem[{Yuan \& Clarke(2004)}]{yuan2004asymptotic}
\textsc{Yuan, A.} \& \textsc{Clarke, B.} (2004).
\newblock Asymptotic normality of the posterior given a statistic.
\newblock \textit{Canadian Journal of Statistics} \textbf{32}, 119--137.

\end{thebibliography}

\end{document}